\documentclass[bibyear]{aa}

\usepackage{natbib}
\usepackage{txfonts}
\usepackage{hyperref}
\usepackage{graphicx}
\usepackage{longtable}

\newcommand*{\factor}{1.0}

\usepackage[svgnames]{xcolor}
\hypersetup{colorlinks,citecolor=Blue}

\def\lum        {\ensuremath{L_\text{1.4~GHz}}}
\def\lsol       {\ensuremath{\text{L}_{\odot}}}
\def\msol       {\ensuremath{\text{M}_{\odot}}}
\def\msolyr     {\ensuremath{\text{M}_{\odot}\,\text{yr}^{-1}}}
\def\msolyrmpc  {\ensuremath{\text{M}_{\odot}\,\text{yr}^{-1}\,\text{Mpc}^{-3}}}
\def\whz        {\ensuremath{\text{W}\,\text{Hz}^{-1}}}
\def\mujybeam {\ensuremath{\mu\text{Jy}\,\text{beam}^{-1}}}
\def\mujy       {\ensuremath{\mu\text{Jy}}}
\def\vmax       {\ensuremath{V_\text{max}}}
\def\qtir       {\ensuremath{q_\text{TIR}}}

\def\totalfa   {\ensuremath{3.16 \pm 0.2}}
\def\totbeta   {\ensuremath{0.32 \pm 0.07}} 

\authorrunning{Novak et al.}
\titlerunning{Cosmic star formation history}

\begin{document}

\title{The VLA-COSMOS 3~GHz Large Project: \\ Cosmic star formation history since $z\sim5$}

\author{
        M.~Novak\inst{1}, 
        V.~Smol\v{c}i\'{c}\inst{1},
        J.~Delhaize\inst{1},
        I.~Delvecchio\inst{1},
        G.~Zamorani\inst{2},
        N.~Baran\inst{1},  
        M.~Bondi\inst{3},
        P.~Capak\inst{4},
        C.~L.~Carilli\inst{5},
        P.~Ciliegi\inst{2},
        F.~Civano\inst{6,7},
        O.~Ilbert\inst{8},
        A.~Karim\inst{9},
        C.~Laigle\inst{10},
        O.~Le~F\`{e}vre\inst{8},
        S.~Marchesi\inst{11},
        H.~McCracken\inst{10},
        O.~Miettinen\inst{1}, 
        M.~Salvato\inst{12},
        M.~Sargent\inst{13},
        E.~Schinnerer\inst{14},
        L.~Tasca\inst{8}
        }

\institute{
Department of Physics, Faculty of Science, University of Zagreb,  Bijeni\v{c}ka cesta 32, 10000 Zagreb, Croatia
\and
INAF - Osservatorio Astronomico di Bologna, Via Piero Gobetti 93/3, I-40129 Bologna, Italy.
\and
Istituto di Radioastronomia di Bologna - INAF, via P. Gobetti, 101, 40129, Bologna, Italy
\and
Spitzer Science Center, 314-6 Caltech, Pasadena, CA 91125, USA
\and 
National Radio Astronomy Observatory, P.O. Box 0, Socorro, NM 87801, USA
\and 
Yale Center for Astronomy and Astrophysics, 260 Whitney Avenue, New Haven, CT 06520, USA
\and 
Harvard-Smithsonian Center for Astrophysics, 60 Garden Street, Cambridge, MA 02138, USA
\and
Aix Marseille Universit\'{e}, CNRS, LAM (Laboratoire d'Astrophysique de Marseille), UMR 7326, 13388, Marseille, France
\and 
Argelander-Institut f\"ur Astronomie, Universit\"at Bonn, Auf dem H\"ugel 71, D-53121 Bonn, Germany
\and 
Institut d'Astrophysique de Paris, UMR7095 CNRS, Universit\;{e} Pierre et Marie Curie, 98 bis Boulevard Arago, 75014, Paris, France
\and 
Department of Physics and Astronomy, Clemson University, Kinard Lab of Physics, Clemson, SC 29634-0978, USA
\and 
Max-Planck-Institut f\"{u}r Extraterrestrische Physik (MPE), Postfach 1312, D-85741 Garching, Germany
\and 
Astronomy Centre, Department of Physics and Astronomy, University of Sussex, Brighton, BN1 9QH, UK
\and 
Max-Planck-Institut f\"{u}r Astronomie, K\"{o}nigstuhl 17, D-69117 Heidelberg, Germany
}

\date{Received ; accepted}

\abstract{
We make use of the deep Karl G. Jansky Very Large Array (VLA) COSMOS radio observations at 3~GHz to infer radio luminosity functions of star-forming galaxies up to redshifts of $z\sim 5$  based on approximately 6\,000 detections with reliable optical counterparts. This is currently the largest radio-selected sample available out to $z\sim5$ across an area of 2~square~degrees with a sensitivity of rms $\approx 2.3~\mujybeam$. By fixing the faint and bright end shape of the radio luminosity function to the local values, we find a strong redshift trend that can be fitted with a pure luminosity evolution $\lum \propto(1+z)^{(\totalfa)-(\totbeta) z}$.
We estimate star formation rates (SFRs) from our radio luminosities using an infrared (IR)-radio correlation that is redshift dependent. By integrating the parametric fits of the evolved luminosity function we calculate the cosmic SFR density (SFRD) history since $z\sim 5$.
Our data suggest that the SFRD history peaks between $2<z<3$ and that the
ultraluminous infrared galaxies (ULIRGs; $100~\msolyr<\text{SFR}<1000~\msolyr$) contribute up to $\sim$25\% to the total SFRD in the same redshift range.  Hyperluminous infrared galaxies (HyLIRGs; $\text{SFR}>1000~\msolyr$) contribute an additional $\lesssim$2\% in the entire observed redshift range.
We find evidence of a potential underestimation of SFRD based on ultraviolet (UV) rest-frame observations of Lyman break galaxies (LBGs) at high redshifts ($z\gtrsim4$) on the order of 15-20\%, owing to appreciable star formation in highly dust-obscured galaxies, which might remain undetected in such UV observations.
}


\maketitle

\section{Introduction}

One of the best methods to follow the buildup of stellar mass through cosmic times relies on inferring the cosmic star formation rate density (SFRD) history \citep[for a  review, see][]{madau14}.
A consensus is achieved regarding recent history, where an exponential decline in SFRD by one order of magnitude from redshift $z\sim2$ to the present day is inferred \citep[e.g.,][]{madau96,haarsma00,hopkins06}. 
On the other hand, with an increasing number of ultra-deep surveys the detection threshold is continually being pushed to higher redshifts (up to $z\sim10$) slowly reaching the epoch of reionization  \citep[e.g.,][]{bouwens14,bouwens15}. The light of the early galaxies is a major factor in the process of reionization \citep[e.g.,][]{bouwens16}, and so accurate SFRD measurements are needed to better understand this epoch.

Although the wealth of observations has increased dramatically in the last decade, we still do not understand the core mechanism that governs star formation rate (SFR) histories of individual galaxies. This is because of our  inability to actually follow these galaxies throughout their evolution. We observe galaxy populations at different cosmic epochs and try to link them in a consistent way. A picture has emerged from this method in which blue star-forming (SF) galaxies evolve into red quiescent galaxies through ways of quenching, such as rapid gas reservoir depletion after major merger interactions or active galactic nuclei (AGN) feedback  \citep[e.g.,][]{bell04,schawinski14}.
On the other hand, \cite{bouche10} presented a quenching-free model based on the cosmological decrease of accretion rates with time, which is able to reproduce the observed SFRD.
Another model has also been proposed that uses simple mathematical lognormal forms for SFRD and individual SFR history to reproduce a wide range of observed relations \citep[e.g.,][]{gladders13,abramson16}. 
When the SFRD history is estimated with sufficient precision it can be used to further constrain semianalytical models of galaxy evolution, thereby deepening our understanding of the underlying physics.

Different SFR tracers can be used over the full electromagnetic spectrum, each with its own benefits and shortcomings \citep[e.g.,][]{kennicutt98}.
The most direct tracer measures ultraviolet (UV) light from young massive stars and can be linked with the amount of star formation in the galaxy \citep[e.g.,][]{buat89}. The rest-frame UV emission is redshifted to optical and infrared (IR) wavelengths for the most distant galaxies; this enables the usage of very sensitive instruments, such as the \textit{Hubble Space Telescope} (HST), to probe this epoch \citep[e.g.,][]{finkelstein15}. Currently, the SFRD in the earliest cosmic times (age of the universe less than 1~Gyr) is constrained almost exclusively with these kinds of observations \citep[see also][]{behroozi13}. 
However, when measuring the rest-frame UV emission one must correct for dust extinction, which drastically diminishes the UV light. Well-constrained attenuation curves are needed to correct for this effect \citep[e.g.,][]{bouwens09}.

When dust grains absorb UV light they re-emit it at IR wavelengths. Therefore, far-IR and sub-mm traces SFR best when the dust content is high, yielding a large optical depth. These observations can suffer from poor resolution and source blending, although this was mitigated with observations with the \textit{Herschel Space Observatory}. Current observations allow IR surveys to constrain the dust content and SFRs up to redshift $z<4$  \citep[e.g.,][]{caputi05,rodighiero10,reddy12,gruppioni13}.
Ultraviolet and IR observations can be combined to obtain a more robust hybrid SFR estimator \citep[e.g.,][]{wuyts11,boquien16}.
With the high-resolution sub-mm window opened by the Atacama Large Millimeter/submillimeter Array (ALMA), these wavelengths can be used to probe dusty submillimeter galaxies (SMGs) and their high star formation rates \citep[e.g.,][]{swinbank14,dunlop16}.

When massive stars undergo supernova explosions, the expanding remnants can accelerate the cosmic ray electrons and give rise to synchrotron radiation, which dominates the radio emission at rest-frame frequencies of $<30$~GHz. The observed nonthermal radio emission offers a dust-unbiased view at sub-arcsecond resolution of star formation processes inside the galaxy, and thus eliminates obscuration, while the high resolution assists counterpart matching \citep[e.g.,][]{seymour08,smolcic09a}.
However, it relies heavily on multiwavelength data to provide galaxy redshift and classification due to the featureless shape of the radio spectrum \citep[e.g][]{condon84}. Furthermore, the SFR calibration for radio luminosities is based on the empirical IR-radio correlation to link nonthermal emission with thermal emission \citep[e.g.,][]{helou85,yun01,bell03}.
This correlation continues to be valid across more than five orders of magnitudes in luminosities and holds at least up to redshift of $z<2$, albeit with some redshift evolution \citep[e.g.,][]{sargent10,magnelli15}, and it is likely to be valid even at earlier times up to $z \lesssim 5$ \citep{delhaize17}.

From observations and evolutionary models, we know that SF galaxies dominate the  faint end of the radio counts \citep[e.g.,][]{condon84,gruppioni03,smolcic08,dezotti10,padovani11a, smolcic17b} and have strongly evolving luminosity functions \citep[see also][]{rowan-robinson93}, therefore deep surveys are needed to probe this population at early cosmic epochs. However, deep surveys have to sacrifice area in order to be feasible, which makes them more susceptible to cosmic over- and underdensities. This cosmic variance can have a strong redshift-dependent impact to any counting statistic employed \citep[e.g.,][]{moster11}.

The Cosmological Evolution Survey (COSMOS) 2 deg$^2$ field \citep{scoville07} is therefore well suited for our studies due to its large area, which should minimize cosmic variance, and excellent multiwavelength coverage, which allows for a precise photometric redshift determination.
With the new Karl G. Jansky Very Large Array (VLA) observations obtained for the VLA-COSMOS 3~GHz Large Project \citep{smolcic17a}, the deepest radio survey to date given the area, we can probe the dust-unbiased SFRD up to redshift of $z\sim5$ with $\sim$6\,000 detections of SF galaxies.
Our radio data best traces high-mass ($M_\star>10^{10}~\msol$) and highly SF galaxies (SFR $>100~\msolyr$), which would also be classified as ultraluminous infrared galaxies (ULIRGs; $L_{\text{TIR, 8-1000~}\mu\text{m}}>10^{12}~\lsol$,  see \citealt{sanders96}). At high redshift, we can also constrain even brighter hyperluminous infrared galaxy (HyLIRG; $L_{\text{TIR, 8-1000~}\mu\text{m}}>10^{13}~\lsol$) populations, which have SFRs that are higher than 1\,000~\msolyr. 
To derive the total SFRD history of the entire radio population in the entire observed redshift range we must rely on extrapolations to lower luminosities below the sensitivity limit.

The paper is organized as follows. In Sect.~\ref{sec:data} we briefly describe the data and selection methods used, which by itself is a topic of an accompanying paper \citep{smolcic17b}. We present methods of constructing luminosity functions and modeling their evolution through cosmic time and our results in Sect.~\ref{sec:lumfun}. The calibration used to to derive SFR from radio luminosities is explained in Sect.~\ref{sec:sfrd} along with the cosmic SFRD history estimated from our data. 
We compare our results to the literature in Sect.~\ref{sec:disc}. Discussion of possible systematics are given in Sect.~\ref{sec:sys}. We finally summarize our findings in Sect.~\ref{sec:summary}.

Throughout the paper we have used the flat concordance Lambda cold dark matter ($\Lambda$CDM) cosmology with the following parameters: Hubble constant $H_0=70$\,km\,s$^{-1}$\,Mpc$^{-1}$, dark energy density $\Omega_\Lambda=0.7$, and matter density $\Omega_\text{m}=0.3$. We assume the \cite{chabrier03} initial mass function (IMF) to calculate SFRs.

\section{Data and star-forming galaxy sample}
\label{sec:data}

The sample of galaxies used in this work is radio selected with ancillary data from the rich multiwavelength coverage of COSMOS, which enables precise determination of redshifts and spectral energy distributions (SEDs).

\subsection{Radio data}
The radio data were obtained with 384 hours of VLA A+C array observations in the S band (2~GHz bandwidth centered around 3~GHz) within the VLA-COSMOS 3~GHz Large Project survey. Details of the observational setup, calibration, imaging, and source extraction can be found in \cite{smolcic17a}. Briefly, 192 pointings were used to obtain a map of the COSMOS 2 square degrees with a uniform $rms$ noise equal to 2.3\,\mujybeam\ and an angular resolution of $0\farcs75$. Imaging was performed using the multiscale multifrequency synthesis \citep{rau11} to ensure good deconvolution of both unresolved and extended sources using the entire available 2~GHz bandwidth at once. Self-calibration of pointings containing brighter sources was performed to improve the fidelity of the image. 
A catalog of source components with a signal to noise (S/N) greater than 5 was extracted using the \textsc{Blobcat} software \citep{hales12}, which relies on a flood fill algorithm to detect contiguous blobs of emission. After visual inspection of multicomponent sources, a final catalog of 10\,830 radio sources was assembled, spanning the entire observed  area of 2.6 square~degrees (approximately 10\,000 radio sources across the central COSMOS two square~degrees). The astrometric accuracy is $0\farcs01$ at the bright end and around $0\farcs1$ for the faintest sources.

\subsection{Optical and near-infrared counterparts}
\label{sec:data_counterparts}
We use the auxiliary data from more than 30 bands in the optical, near-infrared (NIR), and near ultraviolet (NUV) available in the COSMOS field from UltraVISTA DR2, Subaru/Hyper-Suprime-Cam, and SPLASH Spitzer legacy program collected in the COSMOS2015 catalog \citep{laigle16}. The catalog contains $\sim$800\,000 sources with reliable photometry across an area of 1.77~deg$^2$ free of stellar contamination. Photometric redshifts were  computed for all sources by SED fitting using the \textsc{LePhare} code \citep{arnouts99,ilbert06} following methods described in \cite{ilbert13}.

The counterpart matching method is fully described in \cite{smolcic17b}, their Section 3, and briefly summarized below. Owing to high sub-arcsecond resolution of both optical and radio data, and the fact that the radio emission is usually linked with massive bright galaxies, a nearest-neighbor counterpart matching scheme was adopted in combination with a false match probability assignment using a well-constructed background model. Optical-NIR counterparts were assigned to radio sources within a 0$\farcs$8 searching radius if they were deemed reliable. Estimates of false match probabilities were drawn from simulations using a background model that takes the $m_{3.6~\mu\text{m}}$ magnitude distribution of radio counterparts into account. It was designed to consider the optical blocking effect, i.e., missing fainter optical-NIR sources in the COSMOS2015 catalog due to nearby presence of a bright radio counterpart. Given these choices, the percentage of spurious matches in the entire radio sample are negligible. Approximately 11\% of radio sources out of 8\,696 positioned in the unmasked optical-NIR area was not assigned a COSMOS2015 counterpart. Half of those have $S/N<6$ in the radio source catalog making them likely candidates for spurious sources. The false detection probability for radio sources can reach up to 24\% for sources with $5<S/N<5.1$, and a total of $\sim3$\% of sources in the radio catalog can be considered spurious (see also \citealt{smolcic17a}). 
If additional optical-NIR counterpart candidates are considered from the $i$-band selected catalog \citep{capak07} and the \emph{Spitzer}/IRAC\footnote{Infrared Array Camera} catalog \citep{sanders07}, then $7.6\%$ of radio sources would remain without a counterpart in the same unmasked area \citep[see also][]{smolcic17b}. 
We limit the optical-NIR counterpart matching to the COSMOS2015 catalog for better consistency with work by \cite{delvecchio17} and \cite{delhaize17}. By taking the  fraction of spurious sources into account, we have an average $\sim8\%$ incompleteness in our counterpart sample. The total number of radio sources with assigned COSMOS2015 counterparts used throughout this paper is 7\,729.

We use spectroscopic redshifts from the internal COSMOS catalog (M.~Salvato et al. in prep.) available  for 35\% of our radio sources. These redshifts were used only if the spectra were flagged as reliable and 90\% of those are located at $z<1.5$. Photometric redshifts were used for the remainder of the sample.
We estimate the accuracy of photometric redshifts of our radio sample by comparing them to the above mentioned spectroscopic catalog and find a median $\Delta z/(1+z_s)=0.01$ at all redshifts, and a 4\% catastrophic failure rate, defined as $\Delta z/(1+z_s)>0.15$. At redshifts $z>1.5$ we find a slightly larger median $\Delta z/(1+z_s)=0.04$ and a catastrophic failure rate of 12\%.
\cite{laigle16} report the photometric redshift normalized median absolute deviation of the entire COSMOS2015 catalog in $3<z<6$ to be $\sigma_{\Delta z (1+z_s)}=0.021$ with a catastrophic failure rate of 13.2\% (see their Table 5).

\subsection{Removing galaxies dominated by AGN in the radio band}
\label{sec:excess}
We are interested in measuring the amount of star formation in galaxies from radio observations, disregarding whether the galaxy is an AGN host or not. We are therefore not interested in removing all AGN host galaxies from our sample, but only those that show clear evidence of radio emission dominated by an AGN.
Unlike IR observations where the photometry can be used to trace a dusty torus in AGN \citep[e.g.,][]{donley12}, radio emission linked to star formation and AGN cannot be disentangled without assuming some correlation with emission at other wavelengths.

In order to quantify AGN contribution in each galaxy of our sample, \cite{delvecchio17} in their Section 3 performed a three-component SED fit using the \textsc{sed3fit} code\footnote{Publicly available at
http://cosmos.astro.caltech.edu/page/other-tools} \citep{berta13}. These fits were performed on the COSMOS2015 photometry using the best redshifts available and take the energy balance between the UV light absorbed by the dust and re-emitted in the IR into account, following the approach adopted in \textsc{Magphys} \citep{dacunha08}. In addition, an AGN component including the continuum disk and the dusty torus emission was added to the fit, seeking a best-fit solution via $\chi^2$
minimization \citep{berta13}. 
From these fits \cite{delvecchio17} estimated IR luminosity arising only from the star formation processes and calculated the IR-based (8-1000~$\mu$m) star formation rate SFR$_\text{IR}$  using the \cite{kennicutt98} relation and the Chabrier IMF. 
These SFR$_\text{IR}$ values should be correlated with radio luminosities \lum\ given the existence of the IR-radio correlation. In order to quantify this correlation and find outliers, these investigators construct histograms of $r=\log(\lum/\text{SFR}_\text{IR})$ in different redshift bins and fit a Gaussian distribution to the histogram (see \citealt{delvecchio17}, Section 4.2). 
The distributions of $r$ peak at higher values with increasing redshift, and in each redshift bin they are skewed toward higher values, corresponding to higher radio luminosities. 
\cite{delvecchio17} define radio-excess sources when $r$ deviates more than $3\sigma$ from the obtained peak of the distribution as a function of redshift, i.e.,

\begin{equation}
r=\log\left(\frac{\lum[\whz]}{\text{SFR}_\text{IR}[\msolyr]}\right)>22\times(1+z)^{0.013}
\label{eq:excess_cut}
.\end{equation}

Such a cut enables us to discriminate 1\,814 (23\%) sources dominated by AGN emission in the radio. If we assume that the peak of the above $r$ distribution at some redshift corresponds to the ideal correlation between the \lum\ and the SFR$_\text{IR}$, and all values above the peak are due to the increasing AGN contribution, then we estimate that the above cut corresponds to at least 80\% of the radio emission due to the radio AGN component. 
The choice of this cutoff is somewhat arbitrary and was chosen as a conservative limit by \cite{delvecchio17} to minimize contamination of their AGN sample by SF galaxies. Radio emission of galaxies below this $3\sigma$ threshold might still be partly contaminated by AGN emission, but not likely dominated by it. Possible biases of this selection criterion are further discussed in Sect.~\ref{sec:sys_agn}.

We consider 5\,915 radio sources without radio excess as our main SF galaxy sample. The redshift distribution of this final SF sample as well as radio-excess sources that were removed are shown in the upper panel of Fig.~\ref{fig:lumz}.

\section{Radio luminosity function of star-forming galaxies}
\label{sec:lumfun}

Radio luminosity functions (LFs) at different cosmic epochs are used to measure the evolution of radio sources, while also providing constraints on galaxy evolution models.
We first discuss methods of determining the LF from our detections, we then show how the data can be fitted with an analytical form and, finally, we present LFs for our SF galaxies up to redshift of $z\sim5$.

\subsection{Estimating the luminosity function from the data}
\label{sec:method_lf}
\begin{figure}
\centering
\includegraphics[ width=\factor\columnwidth]{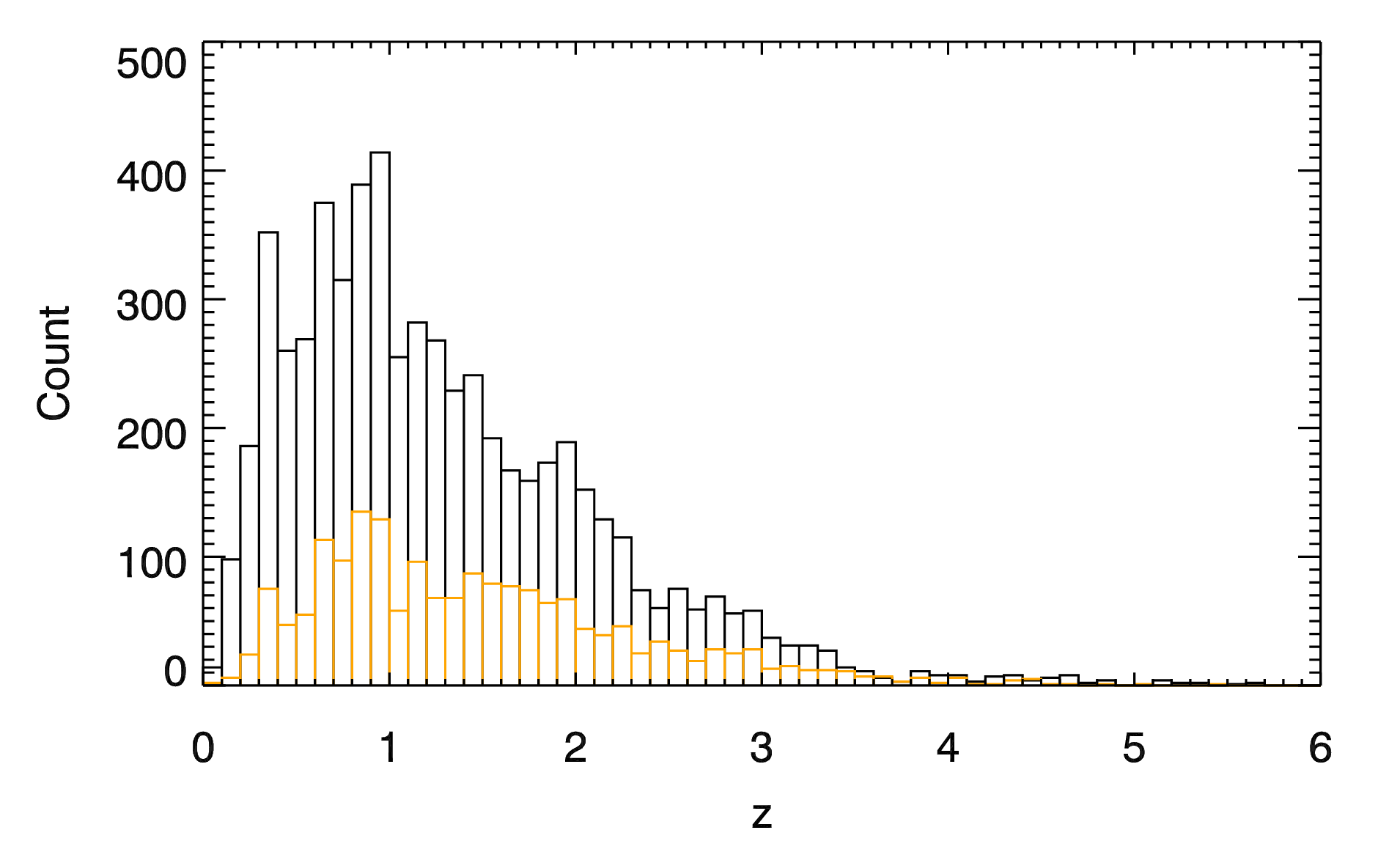}

\includegraphics[ width=\factor\columnwidth]{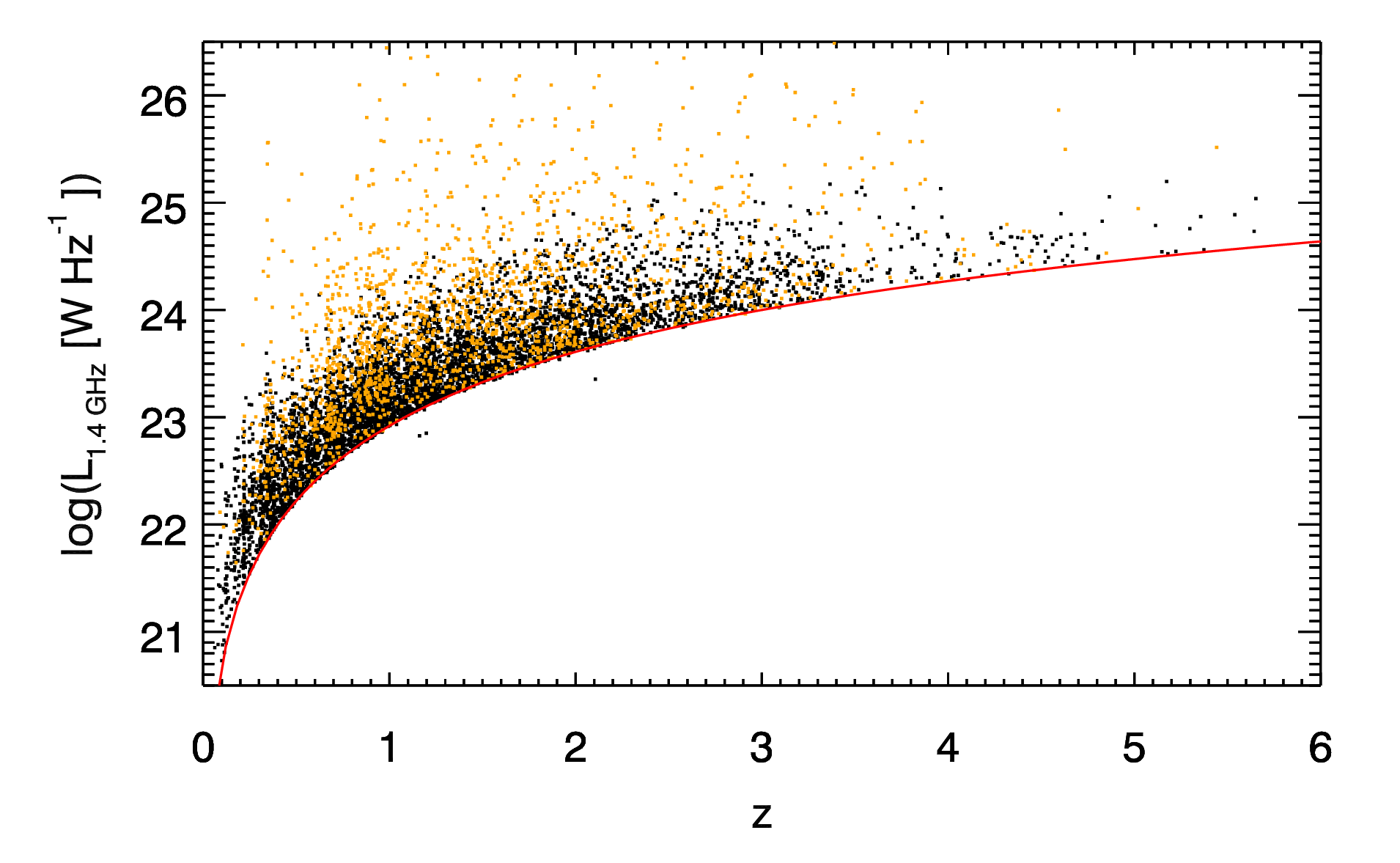}
\caption{Number (\textit{top}) and rest-frame 1.4~GHz luminosity (\textit{bottom}) distribution of our SF (black) and radio-excess ANG (orange) galaxies as a function of redshift. The red line indicates the detection limit of $5\sigma$, where $\sigma=2.3~$\mujybeam at 3~GHz and a fixed spectral index of $\alpha=-0.7$ is assumed.}
\label{fig:lumz}
\end{figure}

Throughout this work we assume that radio sources exhibit a radio spectrum described as a simple power law $S_\nu\propto\nu^\alpha$, where $S_\nu$ is a monochromatic flux density at frequency $\nu$ and $\alpha$ is the spectral index.
This leads to the standard radio K correction of $K(z)=(1+z)^{-(1+\alpha)}$. The final expression for the rest-frame radio luminosity $L_{\nu_1}$ at frequency $\nu_1$ derived from the observed flux density $S_{\nu_2}$ at frequency $\nu_2$, redshift $z$, and luminosity distance $D_L$ is, therefore,
\begin{equation}
L_{\nu_1}=\frac{4\pi D_L^2(z)}{(1+z)^{1+\alpha}} \left(\frac{\nu_1}{\nu_2}\right)^\alpha S_{\nu_2} .
\end{equation}
Luminosities calculated at the rest-frame 1.4~GHz as a function of redshift are shown in the bottom panel of Fig.~\ref{fig:lumz}. This frequency is chosen to simplify a comparison of our results with the literature, where 1.4~GHz observations are more common.
For about $\sim$25\% of the sources we were able to derive the spectral index between 1.4~GHz \citep{schinnerer10} and 3~GHz, while for the remaining sources we assumed the standard $\alpha=-0.7$, which is a valid median value for SF galaxies to be expected for shock-accelerated cosmic ray electrons.

To compute the density of sources and subsequently the LF at different cosmic times (i.e., redshift bins), we employed the  $V_\text{max}$ method \citep{schmidt68}. This method uses the maximum observable volume of each source, while satisfying all selection criteria; it is not dependent on the shape of the LF and therefore reduces the sample and selection biases. The LF $\Phi(L,z)$ gives the number of radio sources in a comoving volume per logarithm of luminosity and is obtained as
\begin{equation}
    \Phi(L,z)=\frac{1}{\Delta\log L}\sum_{i=1}^N\frac{1}{V_{\text{max},i}},
\end{equation}
where \vmax$_{,i}$\ is the maximum observable volume of the $i$-th source, $\Delta\log L$ is the width of the luminosity bin, and the sum goes over each source $i$ in a given redshift and luminosity bin. To take into account different effects and biases, such as a luminosity limited sample or nonuniform noise in the radio map, which may lead to an incompleteness of the sample, we employed a very general form for calculating the maximum observable volume $V_\text{max}$, i.e.,
\begin{equation}
V_{\text{max}, i}=\sum_{z=z_{\text{min}}}^{z_{\text{max}}} [V(z+\Delta z) - V(z)] C(z) ,
\label{eq:vmax}
\end{equation}
where the sum starts at the beginning of a chosen redshift bin and adds together comoving volume spherical shells $\Delta V=V(z+\Delta z) - V(z)$ in small redshift steps $\Delta z=0.01$ until the end of the redshift bin is reached.
The parameter $C(z)$ is the  redshift-dependent geometrical and statistical correction factor that takes the observed area  and sensitivity limit into account and further mitigates some of the other already mentioned completeness issues
\begin{equation}
C(z)= \frac{A_{\text{obs}}}{41253~\text{deg}^2} \times C_{\text{radio}}[S_{3~\text{GHz}}(z)] \times C_{\text{opt}}(z),
\end{equation}
where ${A_\text{obs}}=1.77~\text{deg}^2$ corresponds to the effective unflagged area observed in the optical to NIR wavelengths, $C_{\text{radio}}$ is the completeness of the radio catalog as a function of the flux density $S_{3~\text{GHz}}$, and $C_{\text{opt}}$ is the completeness owing to radio sources without assigned optical-NIR counterpart.
The area observed in the radio encompasses the entire area observed in the optical-NIR and does not have flagged (cropped) regions, therefore NIR observations set the limit for the observed area.
The $C_{\text{radio}}$ factor depends on the redshift because a source with a given intrinsic luminosity changes its apparent flux density between $z_{\text{min}}$ and $z_{\text{max}}$ in $V_{\text{max}}$ calculations (see Eq.~(\ref{eq:vmax})).

\begin{figure}
\centering
\includegraphics[ width=\factor\columnwidth]{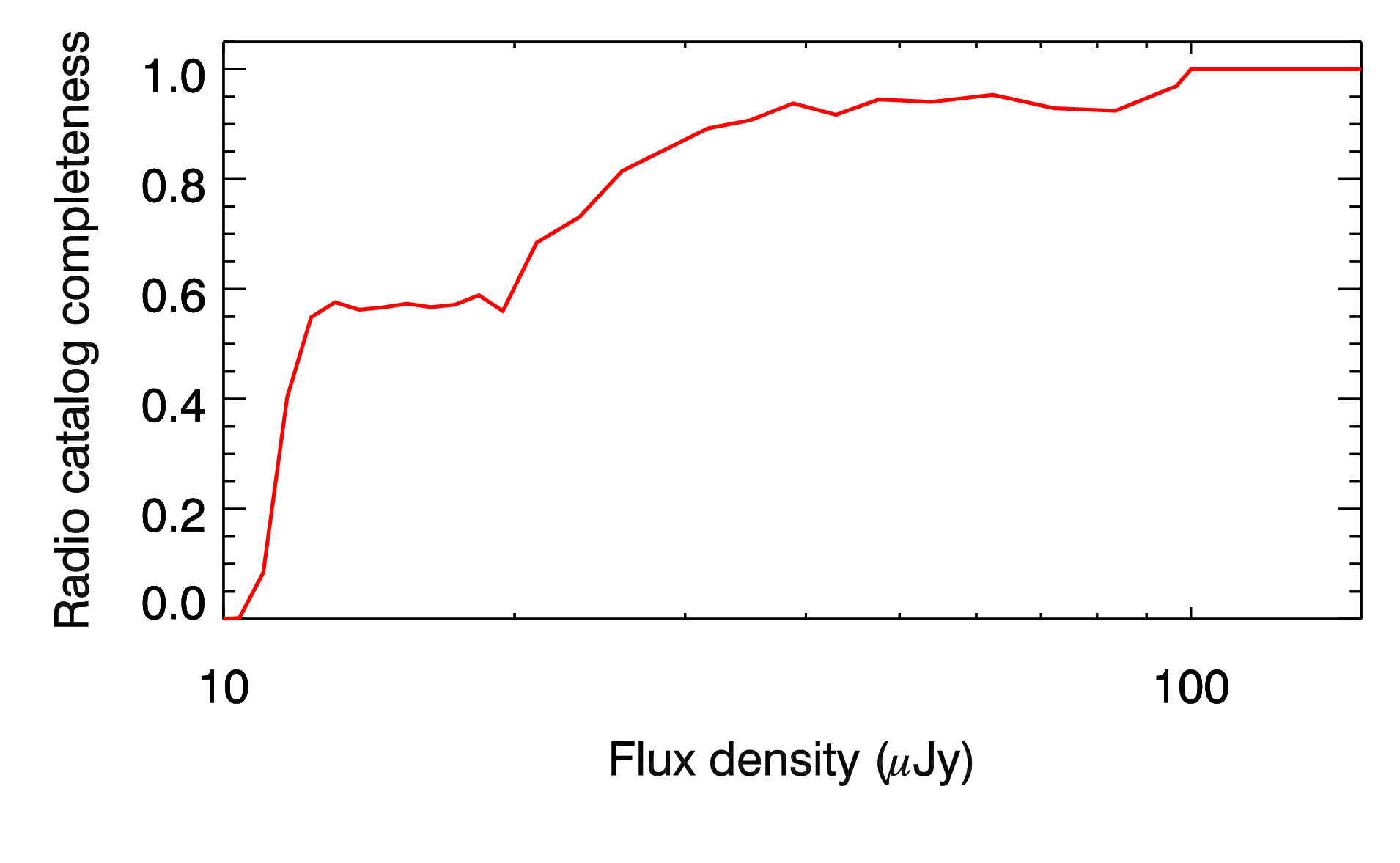}

\includegraphics[ width=\factor\columnwidth]{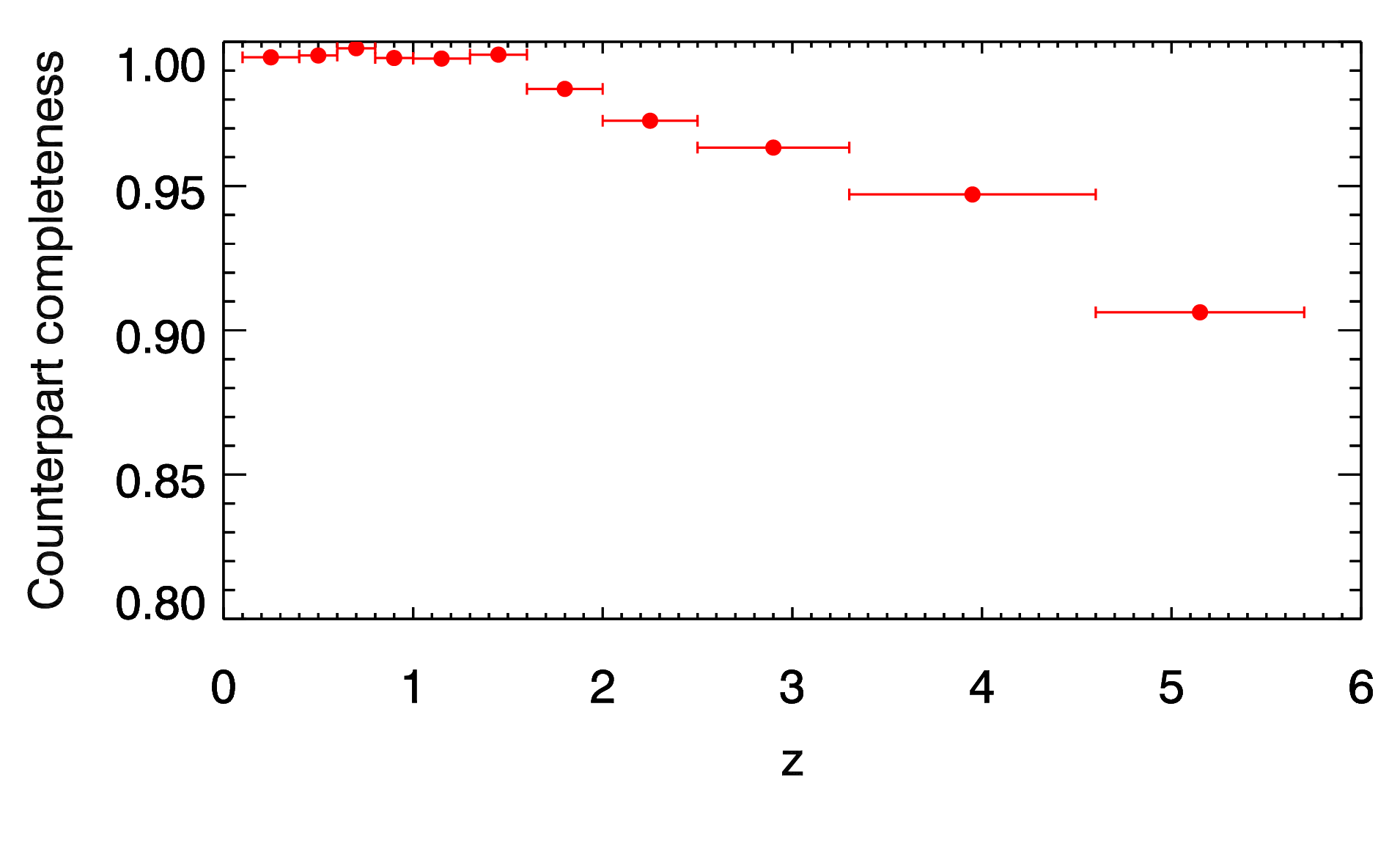}
\caption{\textit{Top}: Radio catalog completeness based on Monte Carlo simulations and mock source insertions \citep[from][]{smolcic17a}. They take into account resolution bias, nonuniform rms, and flux density redistribution due to the source extraction process. \textit{Bottom}: Optical-NIR counterpart completeness based on the amount of additional $i$-band sources that could be matched to our radio sources (see text for details).}
\label{fig:comp}
\end{figure}

Completeness corrections are shown and tabulated in \cite{smolcic17a}; see their Fig.~16 and Table~2, respectively. We also show these corrections in the top panel of Fig.~\ref{fig:comp} in this work.  These completeness corrections are linearly interpolated between tabulated values for any flux density below $S_\text{3~GHz}<100~\mu$Jy. Simulations were not optimized to probe sources with flux densities above $100~\mu$Jy and we assume a 100\% completeness for such sources. Completeness corrections are based on Monte Carlo simulations of mock-source generation and extraction and take into account the nonuniform $rms$, proper derivation of flux densities for low S/N sources and the resolution bias (out-resolving and losing extended low-surface brightness radio emission). The last part, which was modeled by assuming the distribution of radio sizes, follows some functional form of flux densities, which reproduces the observed data (for details see \citealt{smolcic17a}).
These corrections are a function of radio flux density only, meaning that all other physical properties are averaged out. For example, the presence of more resolved and extended (and also low-surface brightness) galaxies at lower redshift as a result of their closer proximity to us, which may introduce a redshift-dependent bias.

In Sect.~\ref{sec:data_counterparts} we mentioned that 11\% of our radio sources were not assigned a counterpart. In the bottom panel of Fig.~\ref{fig:comp} we show the completeness of our radio catalog $C_{\text{opt}}$ due to matching with the COSMOS2015 catalog as a function of redshift. 
It was obtained by considering an additional optical catalog selected in the $i$ band  \citep{capak07}. The counterpart completeness was calculated as $1-N_{i-\text{band}}/N_{i-\text{band or COSMOS2015}}$, where $N_{i-\text{band}}$ is the number of new counterparts assigned only to $i$-band selected sources (1\% of the total radio sample) and $N_{i-\text{band or COSMOS2015}}$ is the number of counterparts assigned to either optical catalog.
As already mentioned, we only use the COSMOS2015 catalog for consistency reasons. Our LFs results are perfectly consistent between themselves whether the actual $i$-band counterparts or the $C_{\text{opt}}$ correction curve is used.
As shown in the bottom panel of Fig.~\ref{fig:comp}, the counterpart sample is complete up to $z\sim 1.5$, and $\sim90\%$ complete at $z\sim5.5$
The addition of this completeness correction, while also considering the 3\% of spurious radio sources, leaves 7\% of real radio sources unaccounted for.
A reliable redshift distribution is not available for these of sources, if the source follows a strong redshift trend, for example, all of these sources are located at $z>3$, it still might bias our high-redshift LFs low.

There might exist a small number of galaxies with a high radio flux density and a faint optical-NIR magnitude whose \vmax\ would be determined by the optical-NIR limit. Since the optical-NIR catalog was selected on the $\chi^2$ image, it does not have a well-defined magnitude limit and therefore we cannot apply a more precise correction on \vmax. This may bias our high redshift LF low as the true \vmax\ would be smaller than what we have used for such sources. However, we do not expect a significant effect since there are only $\sim 10$ sources in our sample that have $K_s$ AB magnitudes fainter than 24.5 where the completeness of the optical-NIR catalog becomes an issue \citep[see][]{laigle16}.

The $rms$ error estimate of the LF in each redshift and luminosity bin is calculated as in \cite{marshall85} by weighting each galaxy by its contribution to the sum
\begin{equation}
   \sigma_\Phi(L,z)=\frac{1}{\Delta\log L}\sqrt{\sum_{i=1}^{N}\frac{1}{V^2_{\text{max},i}} }.
\end{equation}
However, if there are ten sources or fewer in a luminosity bin we used the tabulated upper and lower 84\% confidence intervals from \cite{gehrels86}. These intervals correspond to Gaussian $1\sigma$ errors so that $\sigma_\Phi=\Phi\times\sigma_N/N$, where $\sigma_N$ is the small-number Poissonian statistical asymmetrical error on the measured number of sources. We do not add photometric uncertainties into the error budget, but the redshift bins are chosen to be large enough to mitigate possible problems of sources falling into wrong bins. An additional contribution to the total error budget may arise from the imperfect radio SED (see also Sect.~\ref{sec:sys_alpha}).

\subsection{Local radio luminosity function and its evolution}

Radio LFs of SF galaxies are usually described by four parameter analytical forms such as the power-law plus lognormal distribution from \cite{saunders90}
\begin{equation}
\Phi_0(L)=\Phi_\star\left(\frac{L}{L_\star}\right)^{1-\alpha} \exp\left[-\frac{1}{2\sigma^2}\log^2\left(1+\frac{L}{L_\star}\right)\right],
\label{eq:lflocal}
\end{equation}
where the $L_\star$ parameter describes the position of the turnover of the distribution, $\Phi_\star$ is used for the normalization, $\alpha$ and $\sigma$ are used to fit the faint and bright ends of the distribution,  respectively.

Our deep COSMOS radio observations are best suited to study the high-luminosity end of the LF, especially at higher redshifts ($z>1$), where our data do not sample the faint end of the LF, but instead cover a large observed volume.
If we are interested in the total amount of light emitted from SF galaxies at any redshift we must assume the shape of the LF that is not constrained by our data. These luminosities can be probed with wide and shallow low-resolution radio surveys of the local universe, such as the NVSS\footnote{National Radio Astronomy Observatory (NRAO) VLA Sky Survey.} \citep{condon98}. There are a number of works related to the calculation of the local radio LF of  SF galaxies \citep[e.g.,][]{condon89,condon02,sadler02,best05,mauch07} and they are all broadly consistent in the luminosity range of $21.5<\log \lum[\whz]<23.5$.

\begin{figure}
\centering
\includegraphics[ width=\factor\columnwidth]{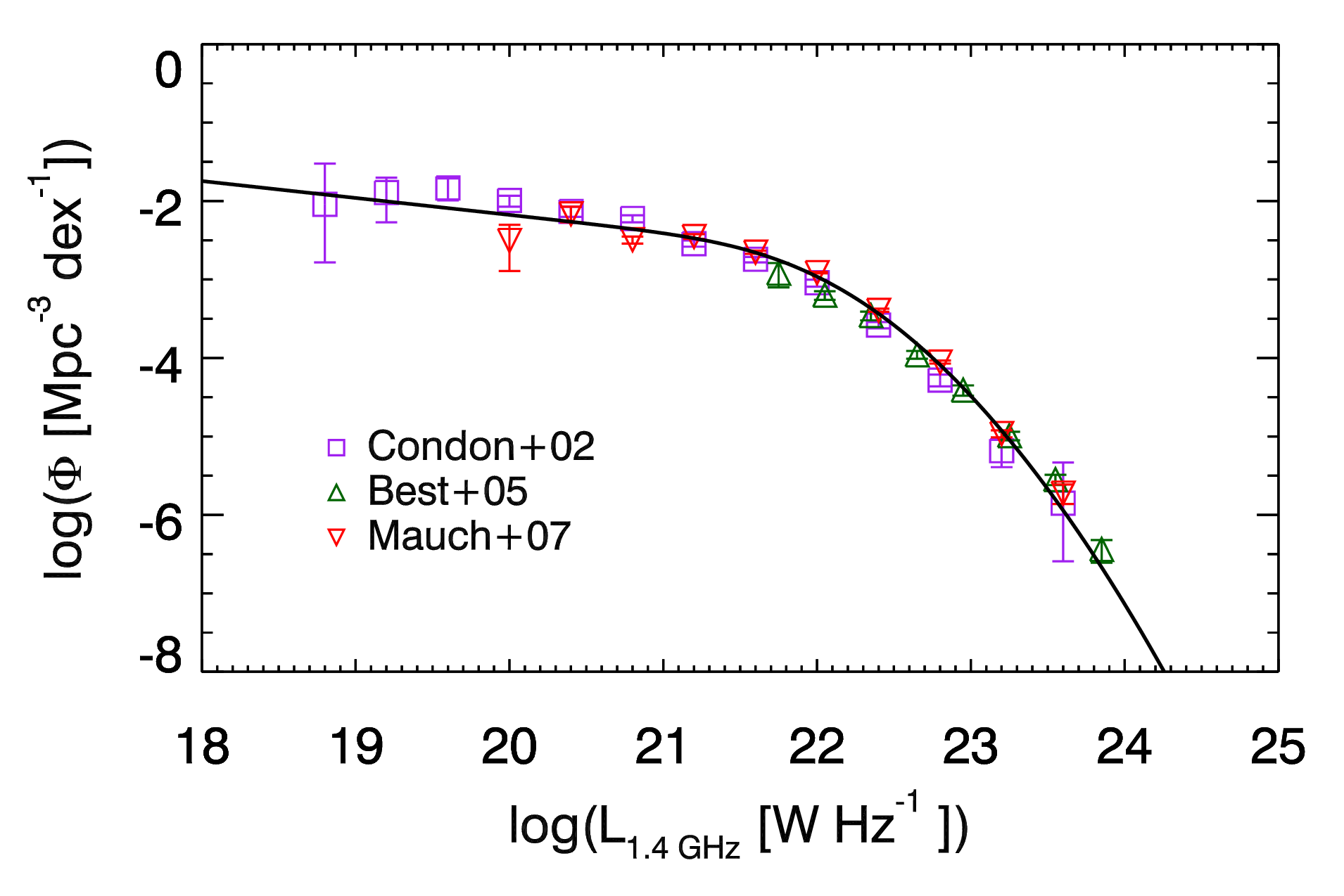}
\caption{
Local radio LF of SF galaxies from several surveys with different observed areas and sensitivities (colored data points) and our fit to the combined data (black line).
}
\label{fig:local_fit}
\end{figure}

To obtain the local luminosity function that is used throughout this work, we performed a fit on combined volume densities from \cite{condon02,best05,mauch07} using the form given in Eq.~(\ref{eq:lflocal}). By combining the data from both wide and deep surveys we can properly constrain both the faint and bright end of the local LF. 
The data and fit are shown in Fig.~\ref{fig:local_fit}. Obtained best-fit parameters are $\Phi_\star=3.55\times10^{-3}~\text{Mpc}^{-3}\text{dex}^{-1}$, $L_\star=1.85\times10^{21}~\whz$, and $\alpha=1.22$, $\sigma=0.63$.

We assume that the shape of the LF remains unchanged at all observed cosmic times and allows only the position of the turnover and normalization to change with redshift. This corresponds to the translation of the local LF in the $\log L - \log \Phi$ plane \citep{condon84} and can be divided into pure luminosity evolution (horizontal shift) and pure density evolution (vertical shift). Using a simple one parameter power law for each of these evolution cases the form of the redshift evolved LF is
\begin{equation}
\Phi(L,z,\alpha_\text{D},\alpha_\text{L})=(1+z)^{\alpha_\text{D}}\Phi_0\left(\frac{L}{(1+z)^{\alpha_\text{L}}}\right),
\label{eq:lumfun_evol}
\end{equation}
where $\alpha_\text{D}$ and $\alpha_\text{L}$ represent pure density  and pure luminosity evolution parameters, respectively, and $\Phi_0(L)$ is given in Eq.~(\ref{eq:lflocal}). Since our data are more sensitive to the most luminous star-forming galaxy population above the knee of the LF, these two evolution parameters may become degenerate preventing a precise estimate of the knee location, especially at higher redshifts. 
This choice for the LF evolution is chosen for its simplicity given that our data constrain the bright end of the LF the best (see also Sect.~\ref{sec:sys_locallf}). In reality, all four parameters may change with redshift.

\subsection{Radio luminosity functions across cosmic times}
\label{sec:res_lumfun}
\begin{figure*}
\centering
\includegraphics[width=\textwidth]{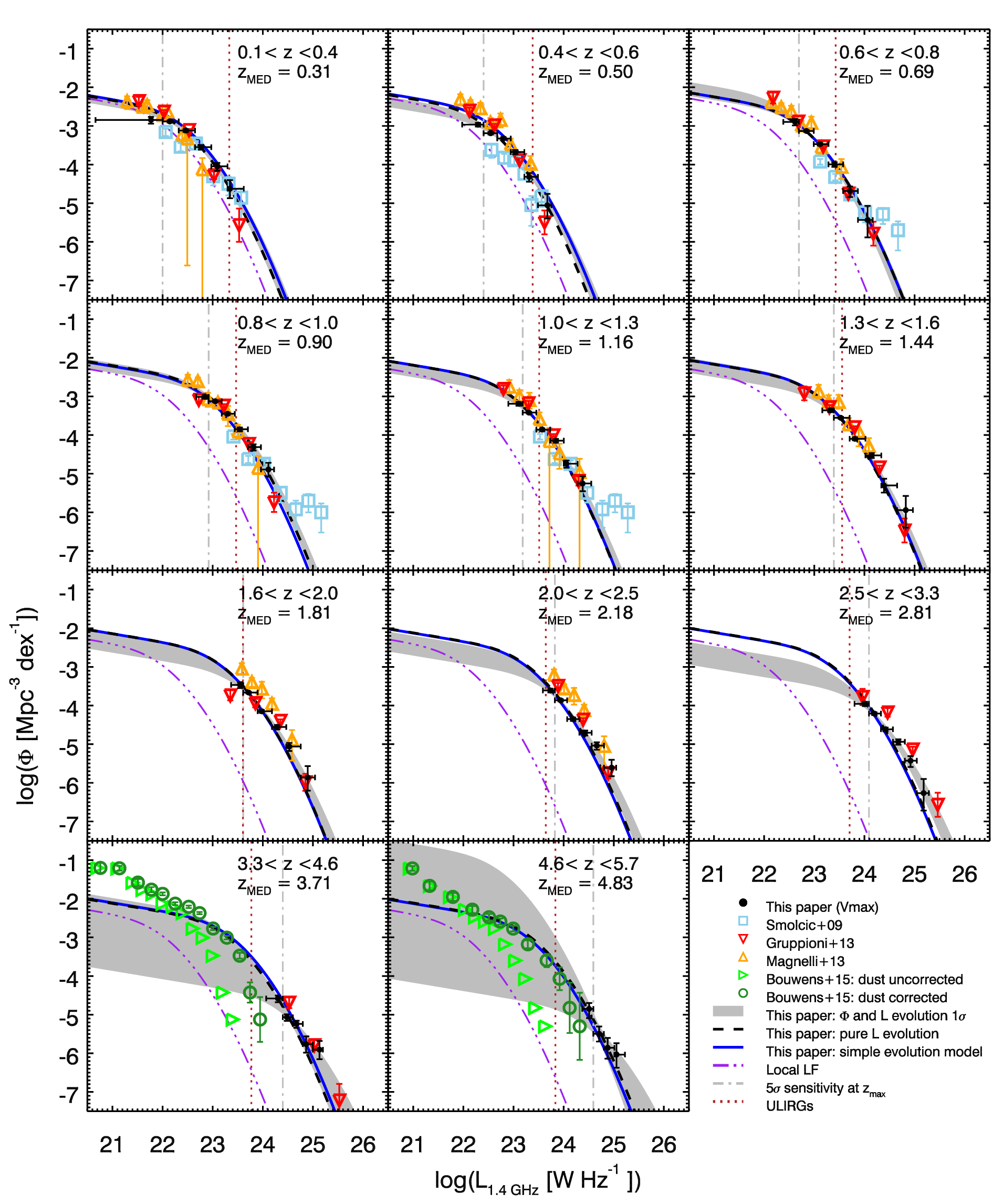}
\caption{Radio luminosity functions of star-forming galaxies in different redshift bins (black filled circles). Best-fit pure luminosity evolved function in each redshift bin is shown with black dashed lines. 
Combined luminosity and density evolution are shown by the gray shaded area (using 68\% confidence region in  $\alpha_\text{D}$, $\alpha_\text{L}$ parameter space around the minimum $\chi^2$).
The local radio function is shown  for reference as a triple-dot-dashed purple line. The vertical dot-dashed line corresponds to the $5\sigma$ luminosity limit at the high redshift end of the bin ($1\sigma=2.3~\mujybeam$ at 3~GHz) under the assumption of a fixed spectral index $\alpha=-0.7$. The vertical red dotted line defines the radio luminosity corresponding to ULIRGs under the assumption of redshift evolving \qtir.
 The redshift range and median redshift of sources in that bin are given in each panel.
All data shown for comparison are indicated in the legend in the bottom right corner; see Sect.~\ref{sec:disc} for details.
}
\label{fig:grid}
\end{figure*}

The procedure of binning sources into luminosities inherently introduces some biases due to averaging and the chosen bin sizes.
To minimize possible completeness issues at the faint luminosity end within a redshift bin, all sources with luminosities below the observational luminosity limit (corresponding to  $5\sigma=11.5~\mujy$ at 3~GHz) at $z_\text{max}$ of the redshift bin were put into single luminosity bin. All sources above this limit were distributed into equally wide luminosity bins spanning the observed luminosity range. The actual luminosity value of each point that we report is the median of all galaxies in a given luminosity bin, while horizontal error bars show the bin width.
For easier comparison with work in the literature, we calculated each LF using the 1.4~GHz rest-frame luminosity obtained from our observations at 3~GHz.
Our LFs from the $V_{\text{max}}$ method are shown in Fig.~\ref{fig:grid} as black circles and are also tabulated in Table~\ref{tab:lumfun_vmax}. 
Our data have small Poissonian error bars due to the relatively large number of sources in each bin and errors do not reflect all possible systematic effects, such as the unknown radio K correction, the error on the completeness, or the sample contamination. A comparison with LFs derived by other authors at different wavelengths is discussed in Sect.~\ref{sec:disc}.

The data points were then fitted with an evolved local LF given in Eq.~(\ref{eq:lumfun_evol}). The redshift that enters this expression is the median redshift of all galaxies in a given redshift bin. 
A $\chi^2$ minimization was performed to obtain the best fit $\alpha_\text{L}$ and $\alpha_\text{D}$ parameters.
Since the LF may have asymmetric errors in sparsely populated bins due to small number Poissonian statistics, an average value of the upper and lower errors on the LF was taken for the $\chi^2$ computation.
These parameters are degenerate when either the faint or bright ends are not sampled well, therefore  a pure luminosity evolution ($\alpha_\text{D}=0$) was computed as well.
Errors on the parameters were estimated from the $\chi^2$ statistics following \cite{avni76}. 
We derived the formal $1\sigma$ errors by projecting onto each parameter axis ($\alpha_\text{L}$ and $\alpha_\text{D}$) the 68\% confidence contour around the minimum $\chi^2$.

\begin{figure}
\centering
\includegraphics[ width=\factor\columnwidth]{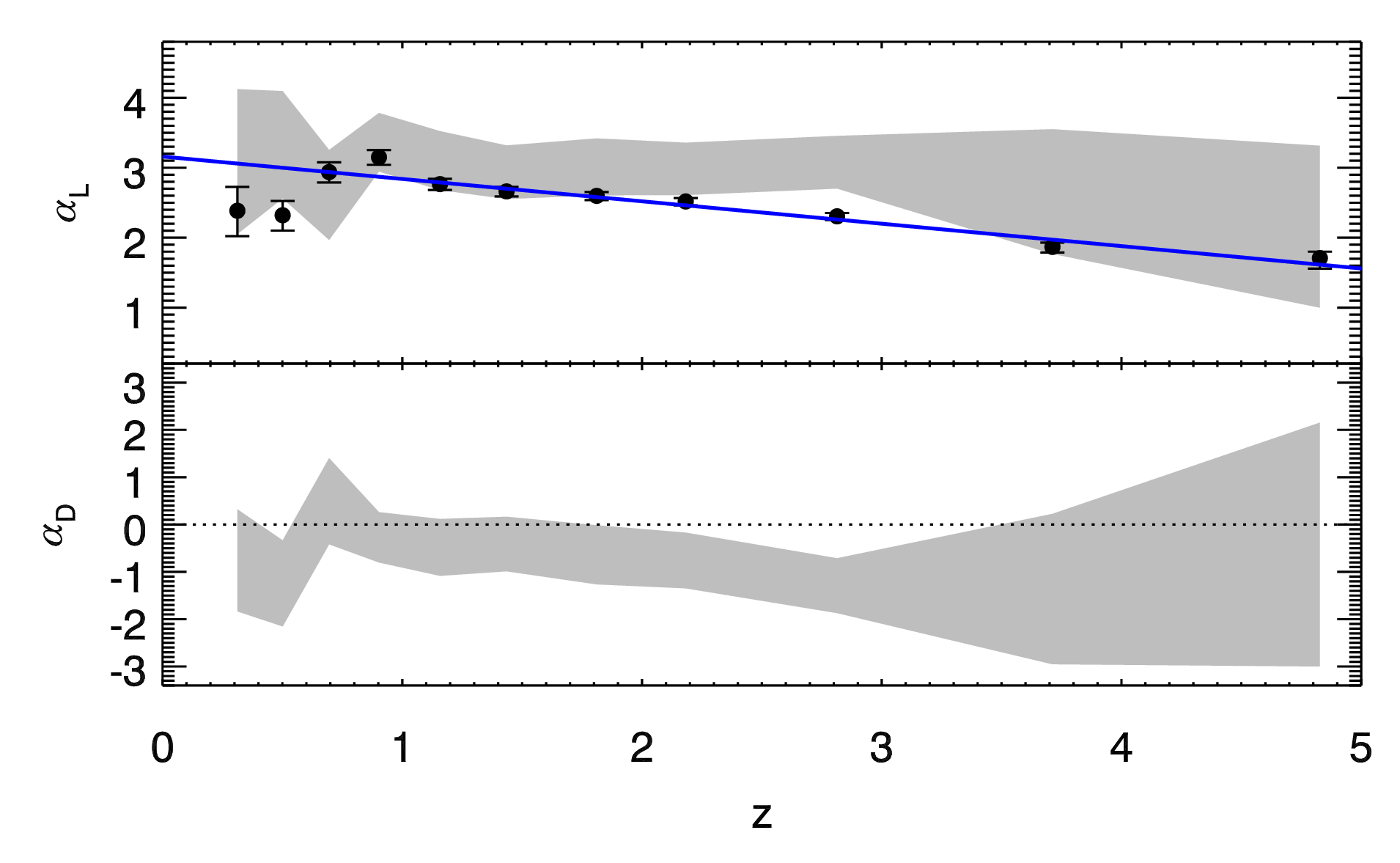}
\caption{Best-fit parameters for the local LF evolution as a function of redshift.  Filled black points correspond to a pure luminosity evolution ($\alpha_\text{D}=0$).
The blue line shows the simple pure luminosity evolution model described in Sect.~\ref{sec:simple_model}.
Gray shaded area shows the 68\% confidence interval for a combined luminosity and density evolution. Large uncertainty in the combined fit is due to parameter degeneracy.
}
\label{fig:knee}
\end{figure}

The best-fit evolution parameters obtained are shown in Fig.~\ref{fig:knee} as a function of redshift.

\subsection{A simple evolution model}
\label{sec:simple_model}

In order to create a single continuous model for the evolution of star-forming LF across the entire observed cosmic time, we simultaneously fit all LF points in all redshift bins 
with a two parameter pure luminosity evolution described as
\begin{equation}
\Phi(L,z,\alpha_\text{L},\beta_L)=\Phi_0\left[\frac{L}{(1+z)^{\alpha_\text{L}+z\beta_L}}\right],
\label{eq:lumfun_evol_tot}
\end{equation}
where $\Phi_0$ is the local LF from Eq.~(\ref{eq:lflocal}), and we allow for an additional redshift-dependent change of the power law parametrized with $\beta_L$. This form follows single redshift bin fits well (see Fig.~\ref{fig:knee}) and is chosen for its simplicity.
Significant density evolution cannot be properly constrained by our observations, which is why we do not attempt it here.
From the $\chi^2$ minimization fit we obtain the following values for parameters: $\alpha_\text{L}=\totalfa$ and $\beta_L=-\totbeta$.

\section{Cosmic star formation rate density history}
\label{sec:sfrd}

\subsection{From radio luminosity to star formation rate}

Radio emission can be used as an extinction-free tracer of star formation rate when linked to other more direct (thermal) tracers such as the IR light.
The first assumption is that the UV photons of massive young stars are absorbed by the dust and re-emitted in the IR so that the total IR emission of a galaxy correlates well with its SFR, which is valid for optically thick galaxies. The conversion factor relies on estimating mass from light and
was calibrated by \cite{kennicutt98} assuming the \cite{salpeter55} initial mass function ($\text{d}N/\text{d}M\propto M^{-2.35}$) from 0.1 to 100~\msol\ and is given by
\begin{equation}
\frac{\text{SFR}}{\msolyr} = 4.5 \times 10^{-37} \frac{L_{\text{TIR}}}{\text{W}},
\end{equation}
where $L_{\text{TIR}}$ contains the total integrated IR luminosity of a galaxy between 8-1000~$\mu$m. This IMF produces more low-mass stars than are supported by observations that favor a turnover below $1~\msol$. Since low-mass stars do not contribute significantly to the total light of the galaxy, only the mass-to-light ratio is changed when the \cite{chabrier03} IMF is adopted instead. This leads to a decrease in SFR by a factor of 1.7 \citep[see also][]{pozzetti07} because of there are fewer low-mass stars created. The calibration itself usually leads to a 0.3~dex scatter on a galaxy basis  \citep[see also][]{condon92,murphy11,kennicutt12}.

Radio observations can trace recent star formation of galaxies, and can trace these observations on timescales of up to 100~Myr \citep{condon92}. 
Estimation of a galaxy SFR from the radio data relies heavily on the observational IR-radio correlation that is known to span at least five orders of magnitudes \citep{helou85, yun01}. The IR-radio correlation links the radio luminosity to the TIR luminosity via the \qtir\ parameter defined as
\begin{equation}
\qtir=\log\left(\frac{L_{\text{TIR}}}{3.75\times 10^{12}\,\text{W}}\right)-\log\left(\frac{\lum}{\whz}\right).
\label{eq:qtir}
\end{equation}
Usually, the \qtir\ parameter is taken to be a constant value derived for local galaxies. However, recent works suggest that the \qtir\ value might change with redshift \citep[e.g.,][]{sargent10,magnelli15}. 
In this paper, we used methods from  \cite{delhaize17} who constrained the median of the \qtir\ as a function of redshift using a doubly censored survival analysis for a joint 3~GHz radio and IR-selected sample. They find a decrease of \qtir\ with redshift that can be parameterized with a simple power law. 
To be self-consistent, we ran the survival analysis on the same SF sample as utilized in this work, while also taking into account limits for IR-detected galaxies without a $5\sigma$ significant radio emission, because in their paper \cite{delhaize17} originally used a different sample selection criteria for excluding AGN. 
The obtained evolution of the IR-radio correlation for our sample can be written as
\begin{equation}
\qtir(z)=(2.78\pm0.02) \times (1+z)^{-0.14\pm0.01}.
\label{eq:qtirz}
\end{equation}
The main idea behind the IR-radio correlation is that a linear relation exists between radio and IR luminosities for SF galaxies. 
There is a possibility that the decreasing \qtir$(z)$ actually mimics some complexities of the radio SED at high redshifts such as varying degrees of free-free contribution and inverse Compton losses. Inverse Compton losses off the cosmic microwave background (CMB) lead to suppression of nonthermal radio continuum emission, which would in turn increase \qtir\ with redshift, but the opposite trend was observed by \cite{delhaize17}.
In the case of a more complicated radio SED, a simple power-law K correction is not a valid assumption anymore.
However, the use of a redshift-dependent \qtir(z) parameter when calculating SFR should account for these intrinsic observational limitations under the assumption of a linear IR-radio correlation as explained in more detail in  Sect.~\ref{sec:sys_irr}.

Finally, the expression that converts radio luminosity into SFR obtained from the steps described above can be written as
\begin{equation}
\frac{\text{SFR}}{\msolyr}  =  f_{\text{IMF}} \times 10^{-24} 10^{\qtir(z)}  \frac{\lum}{\whz}, 
\label{eq:sfr}
\end{equation}
where $f_{\text{IMF}}=1$ for a Chabrier IMF and $f_{\text{IMF}}=1.7$ for a  Salpeter IMF.

\subsection{Star formation rate density across cosmic times}
\label{sec:sfrd_ranges}

\begin{figure*}
\centering
\includegraphics[width=\factor\columnwidth]{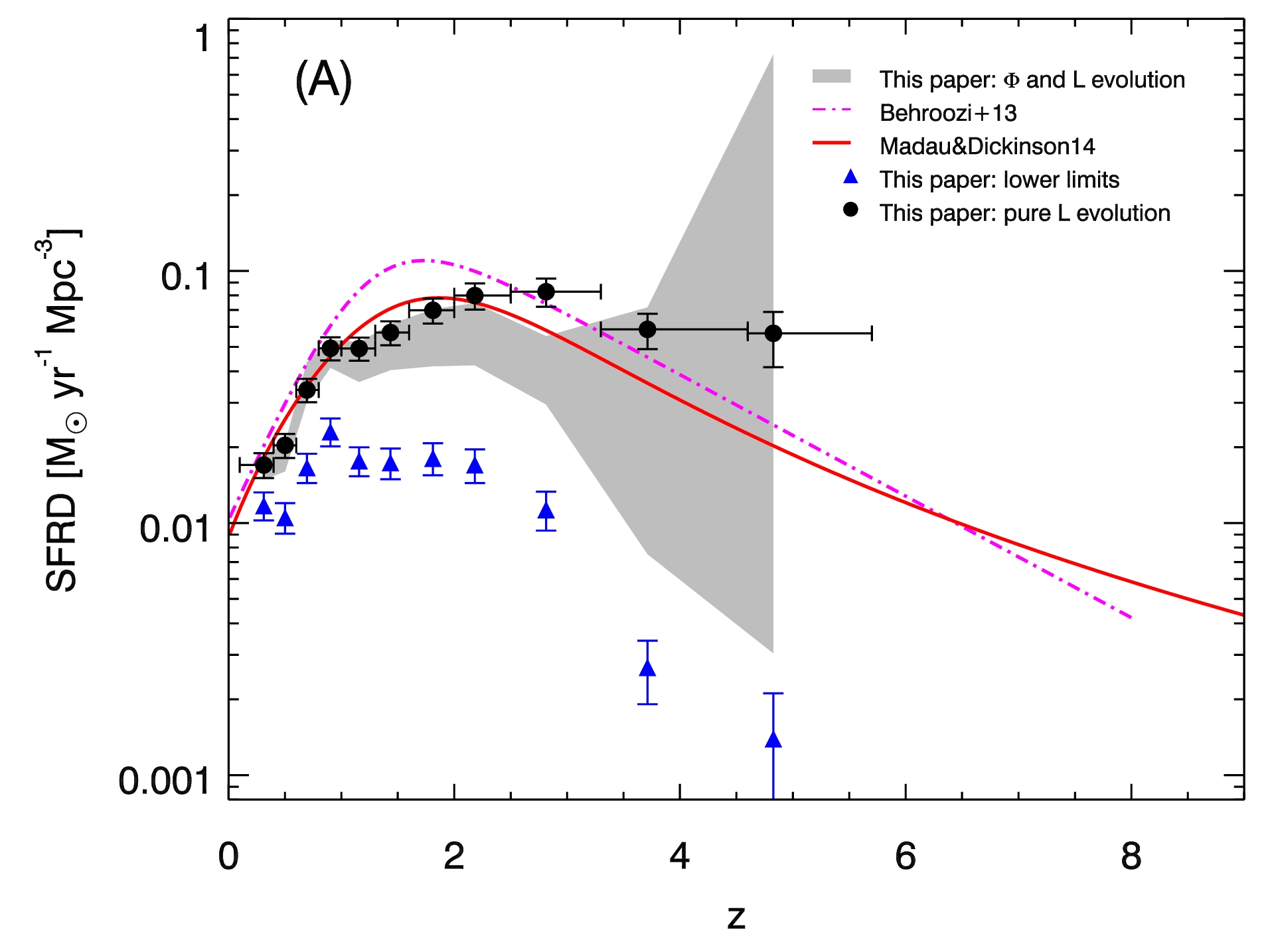}
\includegraphics[width=\factor\columnwidth]{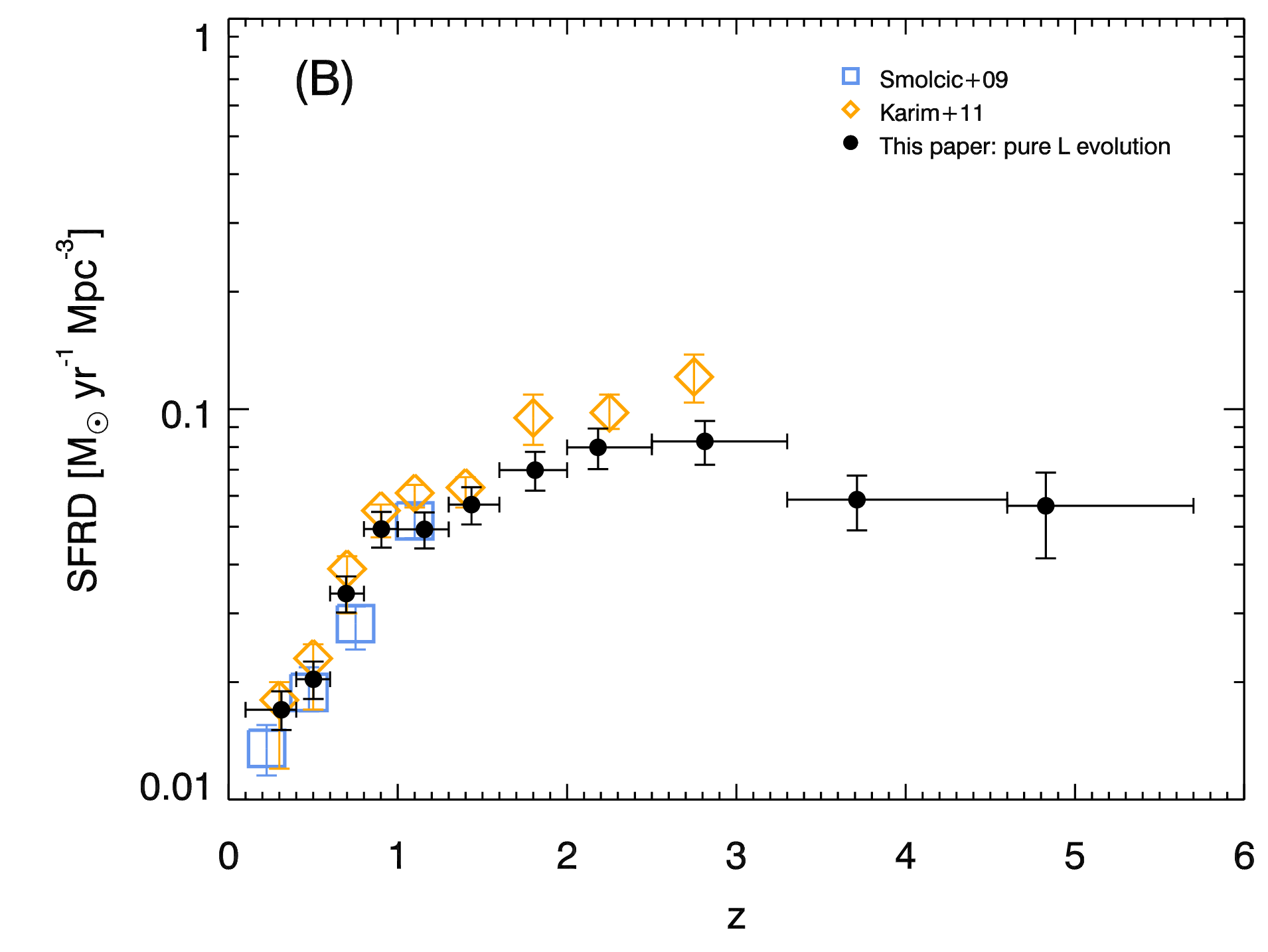}

\includegraphics[width=\factor\columnwidth]{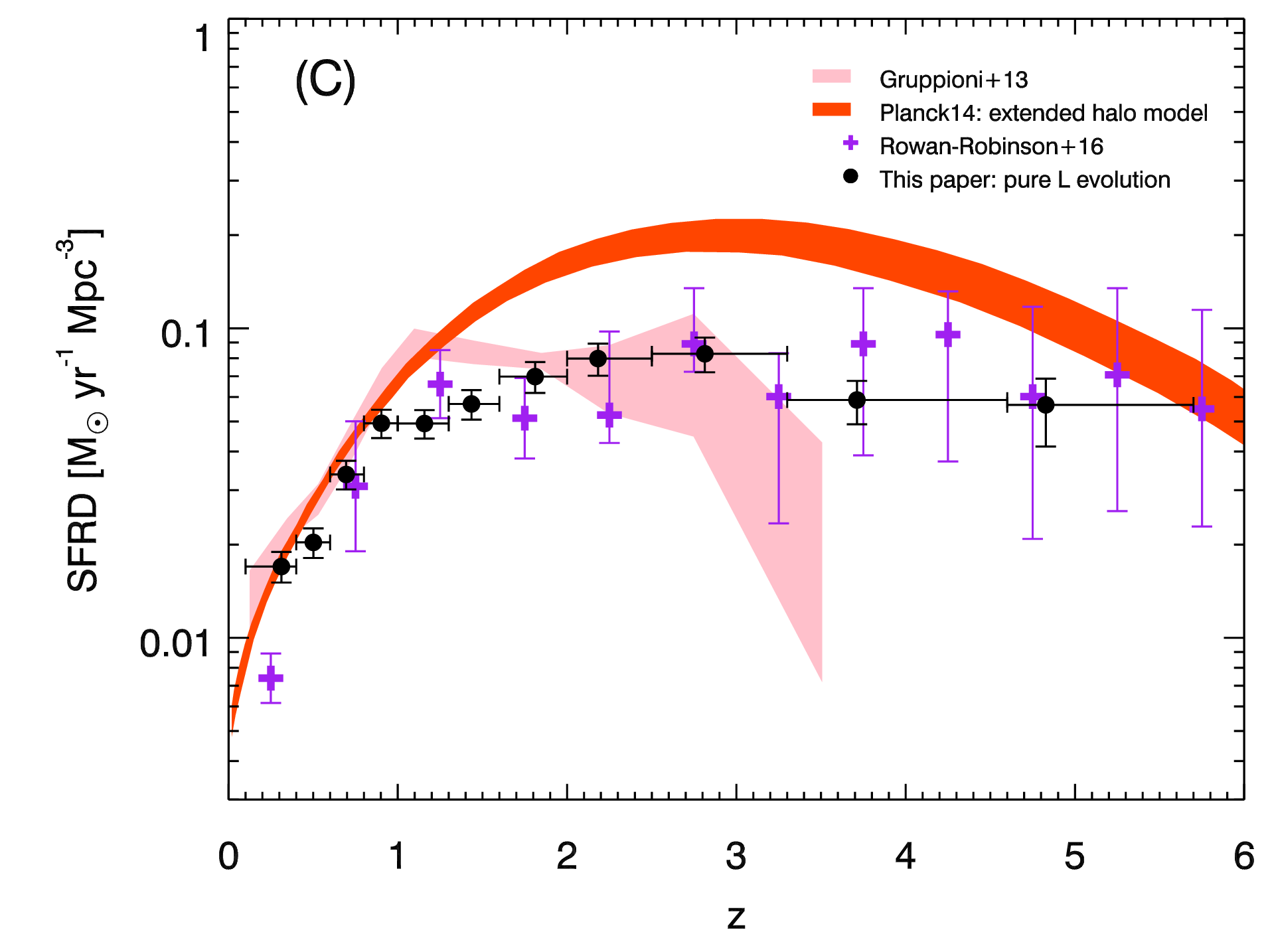}
\includegraphics[width=\factor\columnwidth]{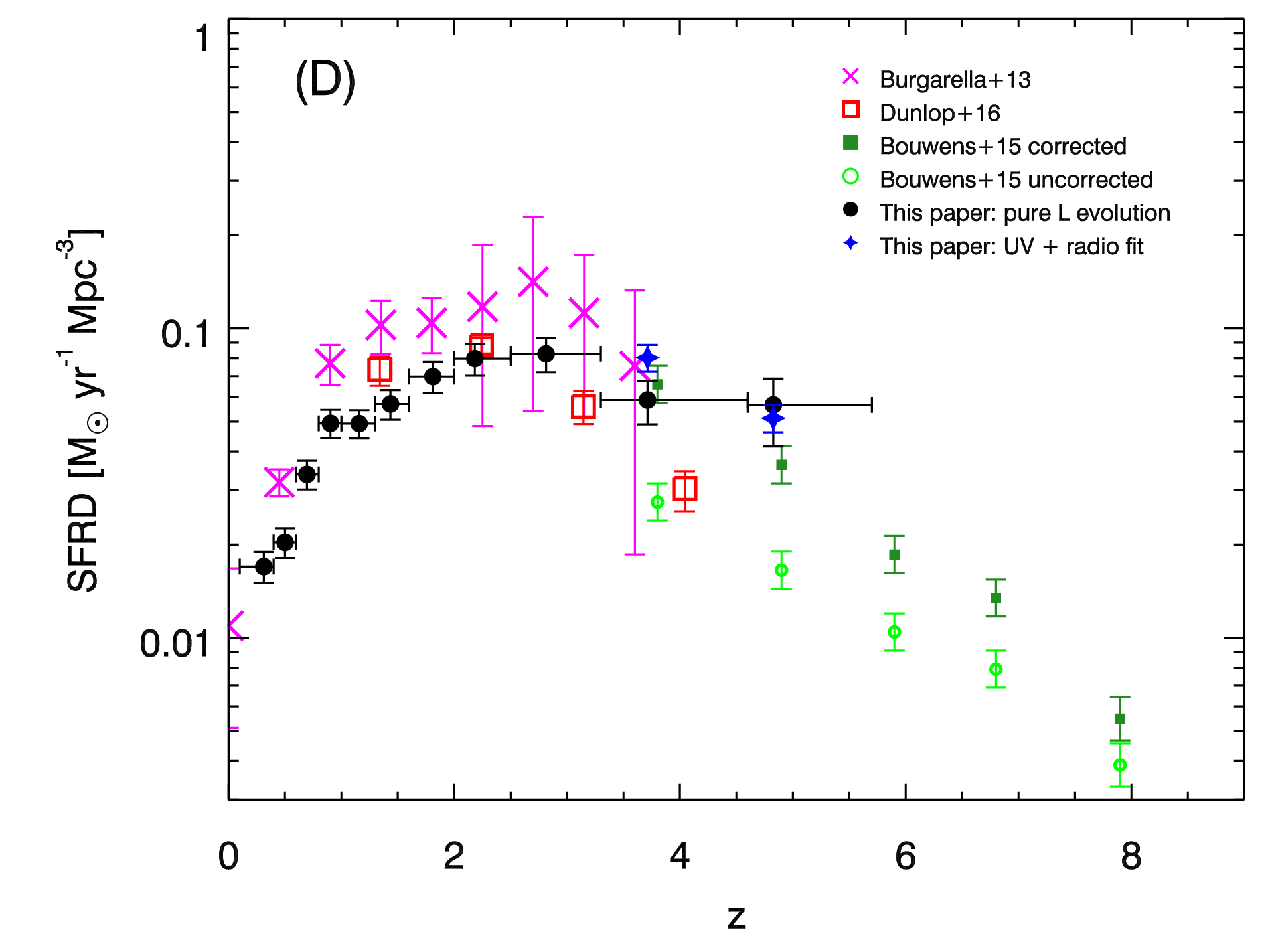}

\includegraphics[width=\factor\columnwidth]{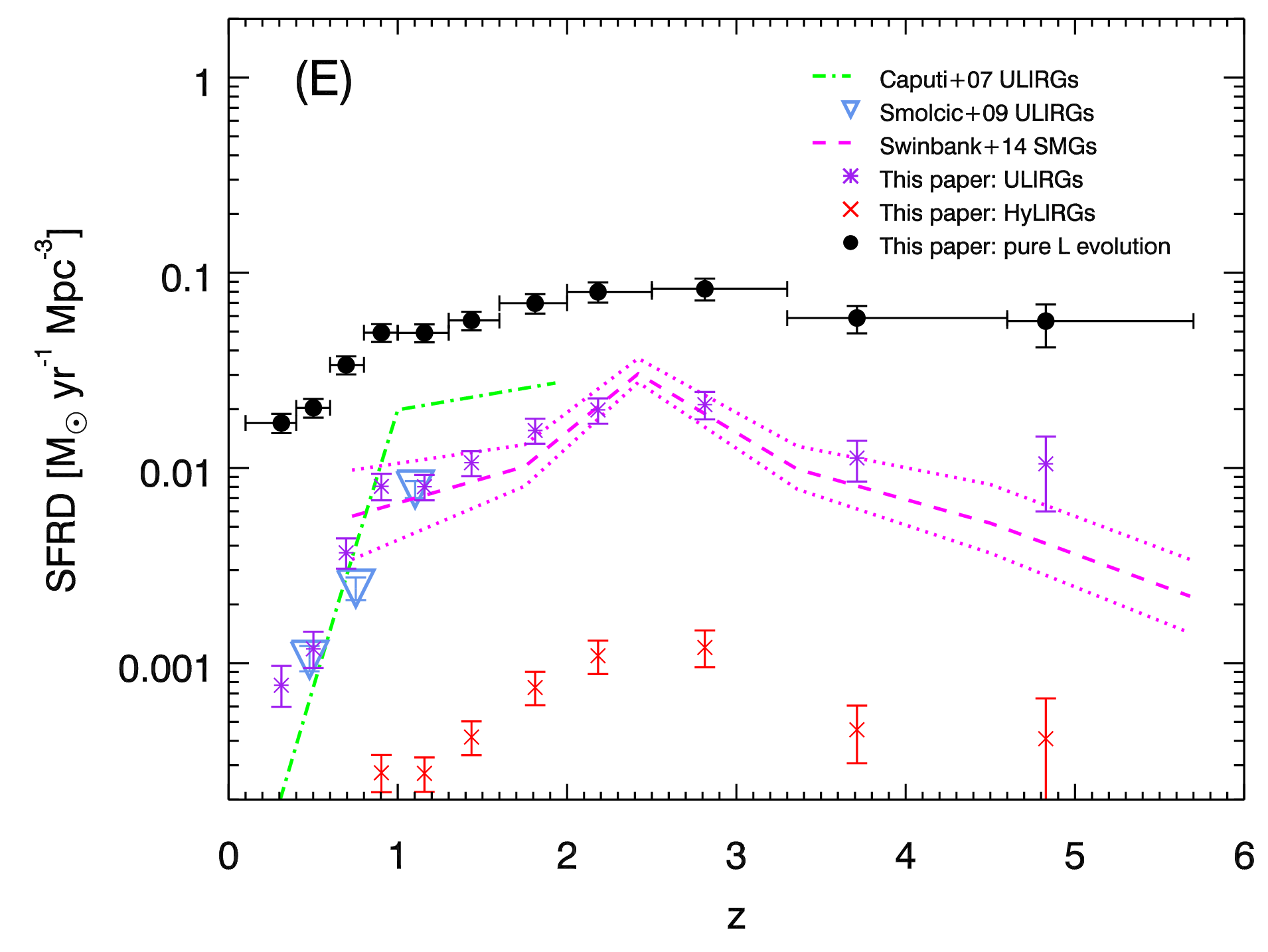}
\includegraphics[width=\factor\columnwidth]{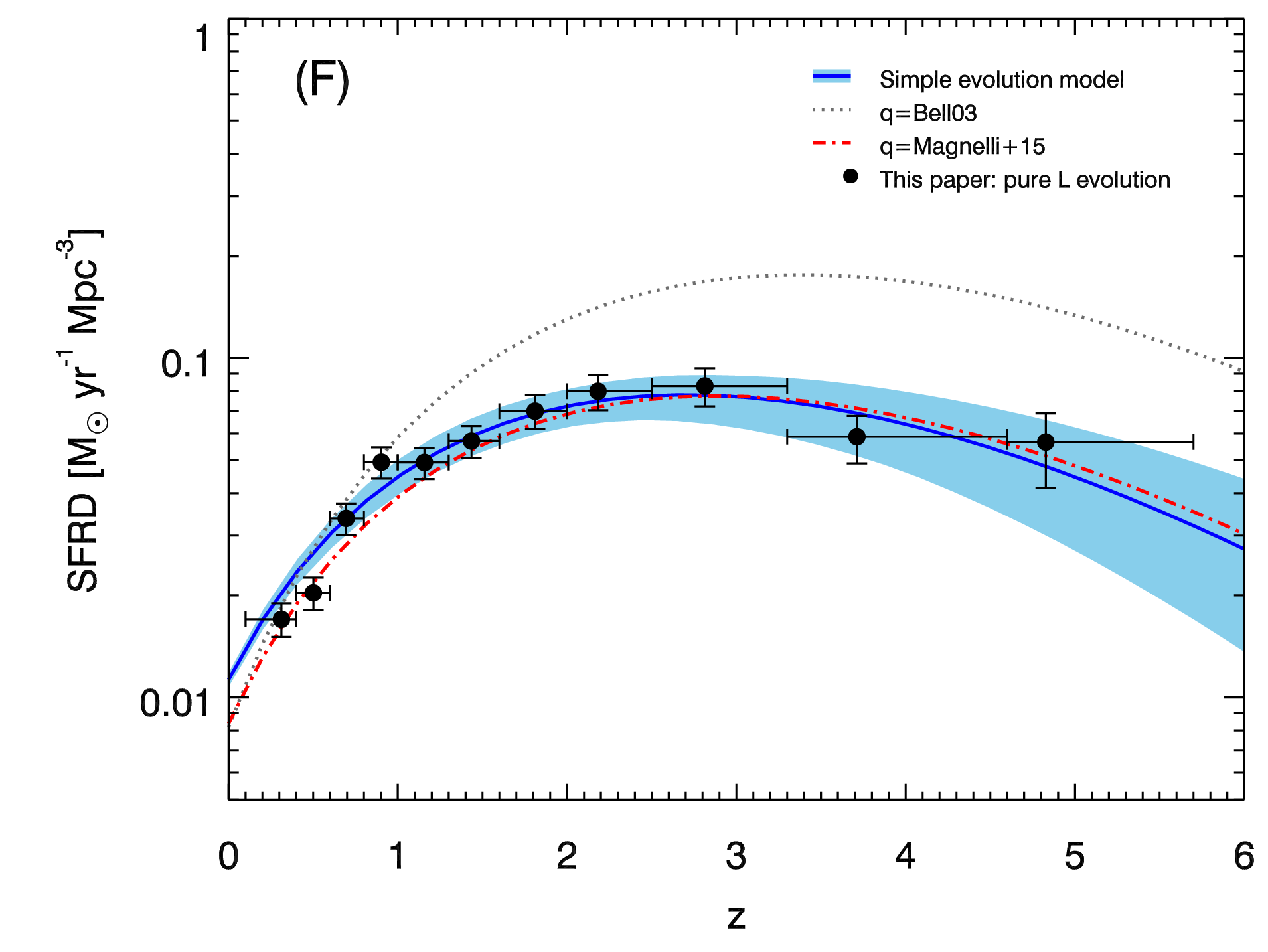}
\caption{Cosmic star formation rate density (SFRD) history. Our total SFRD values estimated from the pure luminosity evolution in separate redshift bins are shown as filled black circles in all panels. All data shown for comparison are indicated in the legend of each panel; see text for details.}
\label{fig:sfrd}
\end{figure*}

Integrating below the LF first multiplied by the luminosity we can obtain the total 1.4~GHz radio luminosity density ($\text{W}~\text{Hz}^{-1}~\text{Mpc}^{-3}$) in a chosen redshift range. Similarly, if the radio luminosity is converted to SFR as given in Eq.~(\ref{eq:sfr}) before integration, the integral yields the SFRD of a given epoch
\begin{equation}
\text{SFRD}=\int_{L_{\text{min}}}^{L_{\text{max}}}\Phi(L,z,\alpha_\text{D},\alpha_\text{L}) \times \text{SFR}(L)\,\text{d}\log L.
\label{eq:sfrd}
\end{equation}
We numerically integrated the above expression by taking the analytical form of the  LF in each redshift bin and using the best-fit evolution parameters shown in Fig.~\ref{fig:knee}. 
The integral was calculated in different luminosity ranges, which are listed below (results shown in Fig.~\ref{fig:sfrd} and also listed in Table~\ref{tab:lumfun_sfrd}):

\begin{enumerate}

\item \textit{Entire luminosity range:} This formally means setting
$L_{\text{min}}=0$ and $L_{\text{max}}\rightarrow+\infty$. The integral converges and the major contribution to the SFRD arises from galaxies with luminosities around the turnover of the LF. The entire radio emission is recovered and if the LF shape and evolution is well constrained the SFRD estimate will be as well (within the SFR calibration errors). This is not the case at higher redshifts ($z>2.5$), where only the bright end of the LF is observed, therefore extrapolation to the faint end can be substantial (see Fig.~\ref{fig:grid}).\\

\item \textit{Data constrained limits:}
$L_{\text{min}}$ and $L_{\text{max}}$ correspond to the lowest and highest value of the observed luminosity function.
By choosing integration limits that correspond to the actual data range, any bias due to LF extrapolation toward higher or lower luminosities is removed. The shape of the local LF also does not affect this result within the fitting errors. Numbers obtained from this integration range are a very conservative lower limit on the SFRD.\\

\item \textit{ULIRGs:}
Limits that correspond to galaxies with IR luminosity of $10^{12}~\lsol<L_{\text{TIR}}<10^{13}~\lsol$ trace ULIRGs. The radio luminosity limits were obtained using an evolving \qtir\ parameter from Eq.~(\ref{eq:qtirz}). The integral with such a range traces SFRD of galaxies that form stars very efficiently (SFR $100-1000~\msolyr$) while also being well constrained by our observational data in $0.5<z<3$ range (see also the red dotted vertical line in Fig.~\ref{fig:grid}).\\

\item \textit{HyLIRGs:}
Similarly, by integrating over galaxies with radio luminosities that translate into $L_{\text{TIR}}>10^{13}~\lsol$, we trace HyLIRGs that have extreme star formation, namely SFR $> 1000~\msolyr$.
\end{enumerate}
Our errors are inferred from the LF fitting parameters uncertainties and added in quadrature with $\qtir(z)$ parameter errors and do not represent the entire error budget due to LF extrapolations.

\section{Comparison with the literature}
\label{sec:disc}

To check the robustness of our LF and SFRD results presented in Figs.~\ref{fig:grid} and \ref{fig:sfrd} and also to create a consistent multiwavelength picture, we compare them with work in the literature derived at radio, IR, UV, and sub-mm wavelengths. All SFR estimates were rescaled to a Chabrier IMF where necessary.

\subsection{Radio and IR luminosity functions}

In Fig.~\ref{fig:grid} we compare our results with the radio LFs by \cite{smolcic09a}, which are based on the VLA-COSMOS 1.4~GHz survey \citep{schinnerer07}. These investigators constructed LFs up to $z<1.3$ using a sample of 340 galaxies classified as star forming using optical rest-frame colors.
The increase in sensitivity of the VLA-COSMOS 3~GHz survey along with a different selection method yielded $\sim10$ times more detections of star-forming galaxies in the same redshift range. The two results generally agree with each other, although our LFs are slightly higher, most likely because of different selection criteria adopted.

We additionally plot LFs from the IR surveys to compare the validity of our results at higher redshifts as well. If the IR-radio correlation is linear, both IR and radio LFs should follow each other well.
To convert the total IR (TIR) to radio luminosity function, the redshift dependent IR-radio correlation parameter \qtir\ described in Eq.~(\ref{eq:qtir}) and (\ref{eq:qtirz}) is used.
We show the LFs by \cite{magnelli13}  derived up to $z<2.3$ from \textit{Herschel} observations of GOODS-N/S deep and GOODS-S ultra-deep fields\footnote{The Great Observatories Origins Deep Survey (GOODS).}. We also show the LFs by \cite{gruppioni13}, which were computed up to $z<4.2$ and are based on \textit{Herschel} PEP/HerMES\footnote{The \textit{Herschel} Guaranteed Time Observation (GTO) PACS Evolutionary Probe (PEP) Survey; The \textit{Herschel} Multi tiered Extragalactic Survey (HerMES).} observations.
To take into account the fact that the redshift bin ranges do not necessarily coincide, we evolved the LFs of other authors using the evolution parameters they reported in their paper, if our median redshift value fell inside their redshift range. Small systematic offsets may arise when the mean redshift does not correspond to the median redshift.
Our data agree well with these surveys both at low and intermediate redshifts. However, at redshift $z>2$ our LFs are systematically slightly lower than those based on IR. Some of this offset might be attributed to a higher percentage of AGN in the IR selected sample at these redshifts \citep{gruppioni13}. They may constitute half of the sample above $z>2.5$.  However, since we start from a differently selected radio sample, exclude AGN identified with radio excess compared to IR emission, and we must rely on the redshift evolving    \qtir$(z)$, it makes the direct comparison difficult.
If a constant $\qtir=2.64$ \citep{bell03} is used for the conversion, instead of an evolving one, our observed radio LFs would actually be higher than  implied by the observed IR-based LFs at high redshifts.

\subsection{UV luminosity functions}

Our radio data are good tracers of highly star-forming and dusty galaxies (ULIRGs and HyLIRGs), but lack the  sensitivity to probe fainter sources at high redshifts. 
We make use of the work performed by \cite{bouwens15} in an attempt to constrain the faint end of the luminosity functions of SF galaxies with actual detections and to simultaneously test their dust corrections.
\cite{bouwens15} utilize \textit{HST}  observations of more than ten deep and wide surveys covering $\sim1000$~arcmin$^2$ to derive the rest-frame UV LFs between $4<z<10$ using a sample of more than 10\,000 Lyman break galaxies (LBGs). The rest-frame UV light correlates strongly with the SFR, unless the galaxy is very dusty. Therefore we can make a broad comparison with our SF galaxy sample.

The SFR calibrations from \cite{kennicutt98} are self-consistent, meaning that all tracers (radio, IR, and UV) should provide the same SFR estimate, thus enabling the link between radio and UV luminosities via the SFR. Although this correlation likely has a large scatter when applied to a specific galaxy, if used on larger samples as a statistical conversion factor, it should allow the conversion of UV magnitudes into radio luminosities. The conversion is needed to compare LFs at these two different wavelengths. The expression for this conversion using the \cite{kennicutt98} calibration, Chabrier IMF, and the redshift-dependent IR-radio correlation from Eq.~(\ref{eq:qtirz}) is
\begin{equation}
\log\frac{\lum}{\whz}=16.556 - 0.4(M_{1600,AB}-A_{UV})-\qtir(z),
\end{equation}
where $M_{1600,AB}$ is the rest-frame UV absolute magnitude reported in \cite{bouwens15} and $A_{UV}$ is the extinction needed to calculate the dust-corrected magnitudes. 
The dust extinction, obtained from the IRX-$\beta$ relationship (correlation between the ratio of FIR to UV fluxes with the UV spectral slope $\beta$; see \citealt{meurer99}), is given in the form of $A_{UV}=4.43-1.99\beta$ and tabulated as a function of UV magnitudes in \cite{bouwens14b}.

The bottom panels in Fig.~\ref{fig:grid} show dust-uncorrected (green right triangles) and dust-corrected (dark green circles) LFs from \cite{bouwens15} for their $z\sim4$ and $z\sim5$ dropouts. Several results can be noted in the plots:

\begin{enumerate}
\item Significant dust corrections are needed at high luminosities and the rest-frame UV cannot be used to detect dustier ULIRGs and HyLIRGs, which are easily observed in the radio at high redshifts. Our radio detections, also available across more than $6\times$ larger area, can therefore provide an independent test for these dust corrections. \\

\item The bright end of the UV LF is lower than our radio LF, with the discrepancy being larger at $z\sim4$ than at $z\sim5$. Our result is also broadly consistent with the result of \cite{heinis13} in which they perform stacking of \textit{Herschel} images of UV selected galaxies at $z\sim1.5$. These authors found that the IR luminosity of bright galaxies ($L_{\text{IR}}>10^{11}~\lsol$) obtained via stacking can recover as low as 52-65\% of the total SFRD derived from the IR-selected samples. \\

\item Even if we disregard the dust correction, the density of faint galaxies two decades in luminosity below our detection limit is very high. The dust-corrected UV LFs, are in broad agreement with our pure-luminosity fit extrapolation more than two decades below the lowest observed radio luminosity at $z\sim5$. These arguments can be used to rule out most of the gray-shaded area in the bottom two panels of Fig.~\ref{fig:grid} arising from significant negative density evolution (see also lower panel of Fig.~\ref{fig:knee}). Rest-frame UV observations are in favor of higher densities of galaxies than what would be obtained if a turnover in the radio LF is introduced immediately below the faintest observed radio luminosity.\\

\item There is a discrepancy between our pure luminosity evolution model and the UV LF at the faintest observed end. Since our radio data cannot constrain such low luminosities, pure luminosity evolution is probably not the best possible model for extrapolating our observed LFs below our detection limits. 
Indeed, a continuous steepening with redshift of the faint end slope of the LF has recently been proposed \citep[e.g.,][]{parsa16}.
\end{enumerate}

We further discuss the UV data in the context of our radio estimates of SFRD in Sect.~\ref{sec:sfrd_uv}.

\subsection{Total SFRD estimates}
\label{sec:sfrd_tot}

Throughout all panels in Fig.~\ref{fig:sfrd} we show our total SFRD derived by integrating the pure luminosity evolved LF in individual redshift bins as black filled circles. 
In panel A we compare our SFRD with the curve from the review by \cite{madau14} where the fit was performed on a collection of previously published UV and IR data (red line).
Below $z<2$ our data agree well with this compilation of data, but show a turnover at higher redshift ($z\sim2.5$) with a shallower decline yielding up to 2-3 times larger SFRD at $z\gtrsim4$.
We also plot a slightly different \cite{behroozi13} fit to the data compilation in the same panel.

If we allow for both  luminosity and  density evolution there is a degeneracy of parameters leading to large uncertainties in the total SFRD estimate; the gray shaded area in panel A of Fig.~\ref{fig:sfrd} is obtained with fit parameters taken from the $1\sigma$ significant region in $\alpha_\text{D}$ and $\alpha_\text{L}$ parameter space. 
We do not fit a pure density evolution because it would increase the normalization of the LF to very high densities. The SFRD estimates would consequently be significantly higher, making our data completely inconsistent with other works in the literature at intermediate and high redshifts.
In the same panel we also show very strict lower limits constrained by the data with blue triangles that demonstrate the amount of extrapolation needed to obtain the total SFRD. 
Although the extrapolation is be significant, especially at higher redshifts, we note
that the UV LFs support the need for such large extrapolations.

\subsection{Comparison with previous radio SFRD}
\label{sec:sfrd_radio}

In panel B of Fig.~\ref{fig:sfrd} we show two radio estimates based on the VLA-COSMOS 20~cm survey \citep{schinnerer07,schinnerer10}. 
\cite{smolcic09a}  calculated the SFRD by integrating the pure luminosity evolution fit of a local LF taken from \cite{sadler02} and their results are shown with blue squares. Also, these estimates were obtained in the COSMOS field and therefore they represent a good consistency check at low redshift. 
A different approach was taken by \cite{karim11} who performed stacking on mass selected galaxies (shown as orange diamonds). They obtained a monotonous rise in the SFRD up to $z\sim3$. Although the field is the same, their method of estimating SFRD is significantly different from ours since it depends on stacking individually undetected sources. Our estimates are slightly lower than theirs, with the difference increasing with redshift. This offset is primarily due to a different IR-radio correlation used. They adopt a calibration from \cite{bell03}, which yields higher SFRD at higher redshifts.

\subsection{Comparison with IR SFRD}
\label{sec:sfrd_ir}

In panel C of Fig.~\ref{fig:sfrd} the pink shaded area shows the $1\sigma$ uncertainty for the SFRD derived from the integrated total IR LF by \cite{gruppioni13}. This LF has a rising trend up to $z\sim1.1$ and then flattens out. The highest redshift estimate should be considered as a lower limit because the PEP selection might miss high-z sources.
Our results are in broad agreement with theirs. Discrepancies at some redshifts might be attributed to different sample selection, since we are excluding AGN host galaxies classified as such only in the radio (see Sect.~\ref{sec:sys_agn}).
Additionally, the agreement between the IR and the radio SFRD is better at $z>2$ than at $z\sim1.5$, while the opposite is true when comparing IR to radio LF (see Fig.~\ref{fig:grid}).
The reason for this effect is that the normalization of the \cite{gruppioni13} IR LF is slightly higher than ours. However, because of the significant negative density evolution and the unchanging faint end slope, this higher normalization is being progressively compensated in their SFRD integral by decreased densities of the faint end. Differing contributions of the faint and the bright end to the total SFRD as a function of redshift lead to apparent agreement between IR and radio SFRD estimates, even though the actual observed LFs do not match perfectly.

We also show the results from the recent work by \cite{rowan-robinson16} as purple plus signs in the plot. Using SED fitting on $\sim3000$ \textit{Herschel} sources from 20.3 square degrees of the sky they derive an IR-based SFRD since $z\sim6$. Even though their result has large uncertainties, the finding supports a much flatter SFRD trend at high redshifts. It is still consistent, however, with our findings  within the error bars.
In the same panel we additionally show SFRD results of an extended halo model estimated by \cite{planck14} from the measured power spectra of the cosmic IR background anisotropies as orange-red shaded area. They report several possible reasons for such high values of SFRD at $z>2$, and it is important to note that all measurements rely on some form of extrapolation toward fainter galaxies. These results might therefore be considered as upper limits.

\subsection{UV addition to SFRD estimates}
\label{sec:sfrd_uv}

In panel D of Fig.~\ref{fig:sfrd} we show the recent rest-frame UV estimates from \cite{bouwens15} as dark-green filled squares (dust-corrected) and green open circles (dust-uncorrected). The SFRD is scaled to Chabrier IMF and \cite{kennicutt98} calibration. Ultraviolet observations like these are well suited to study the early universe owing to the ability to probe exceptionally high redshifts $z\sim10$, as also reviewed in \cite{madau14}.

\begin{figure}
\centering
\includegraphics[ width=\factor\columnwidth]{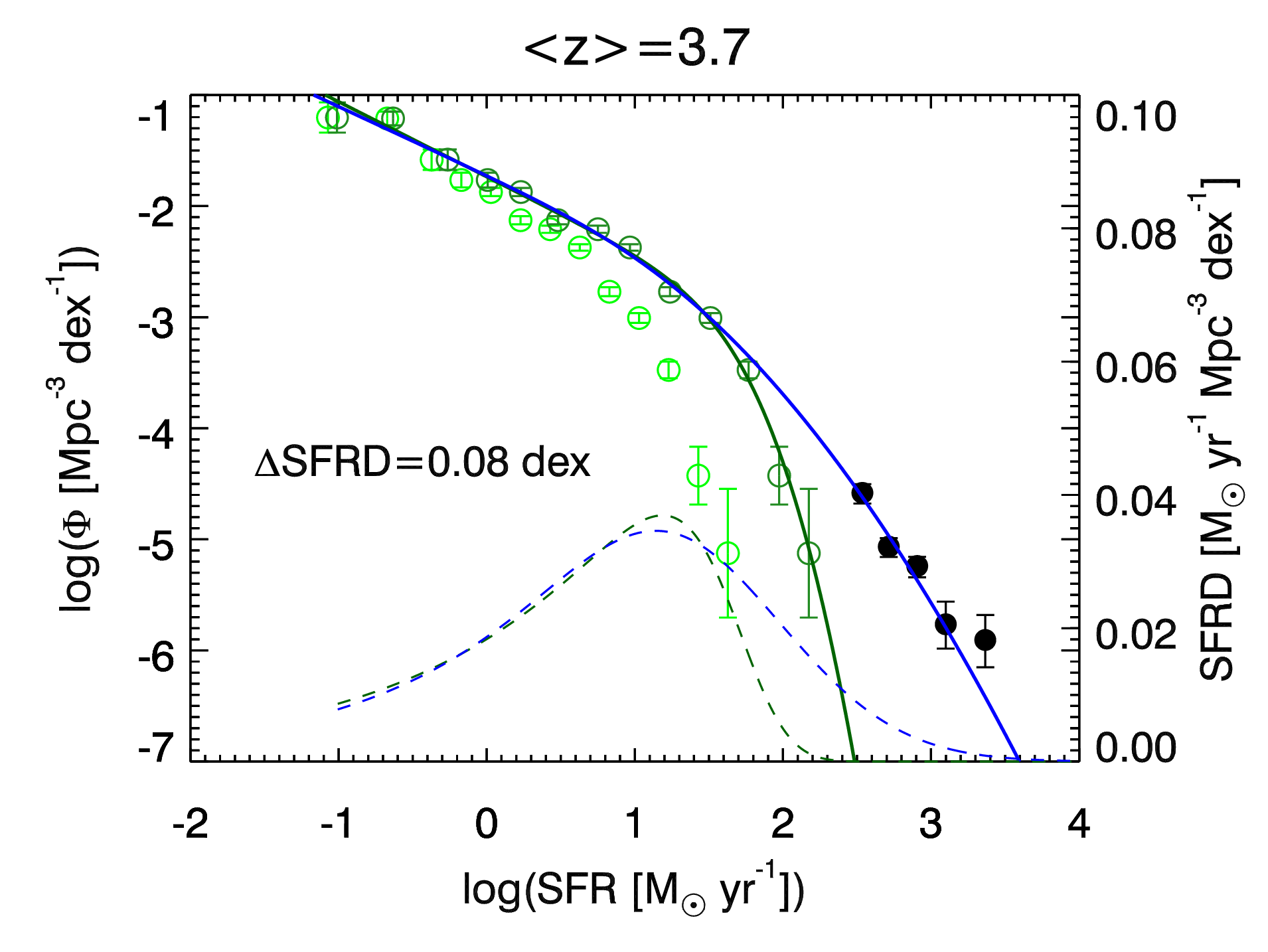}

\includegraphics[ width=\factor\columnwidth]{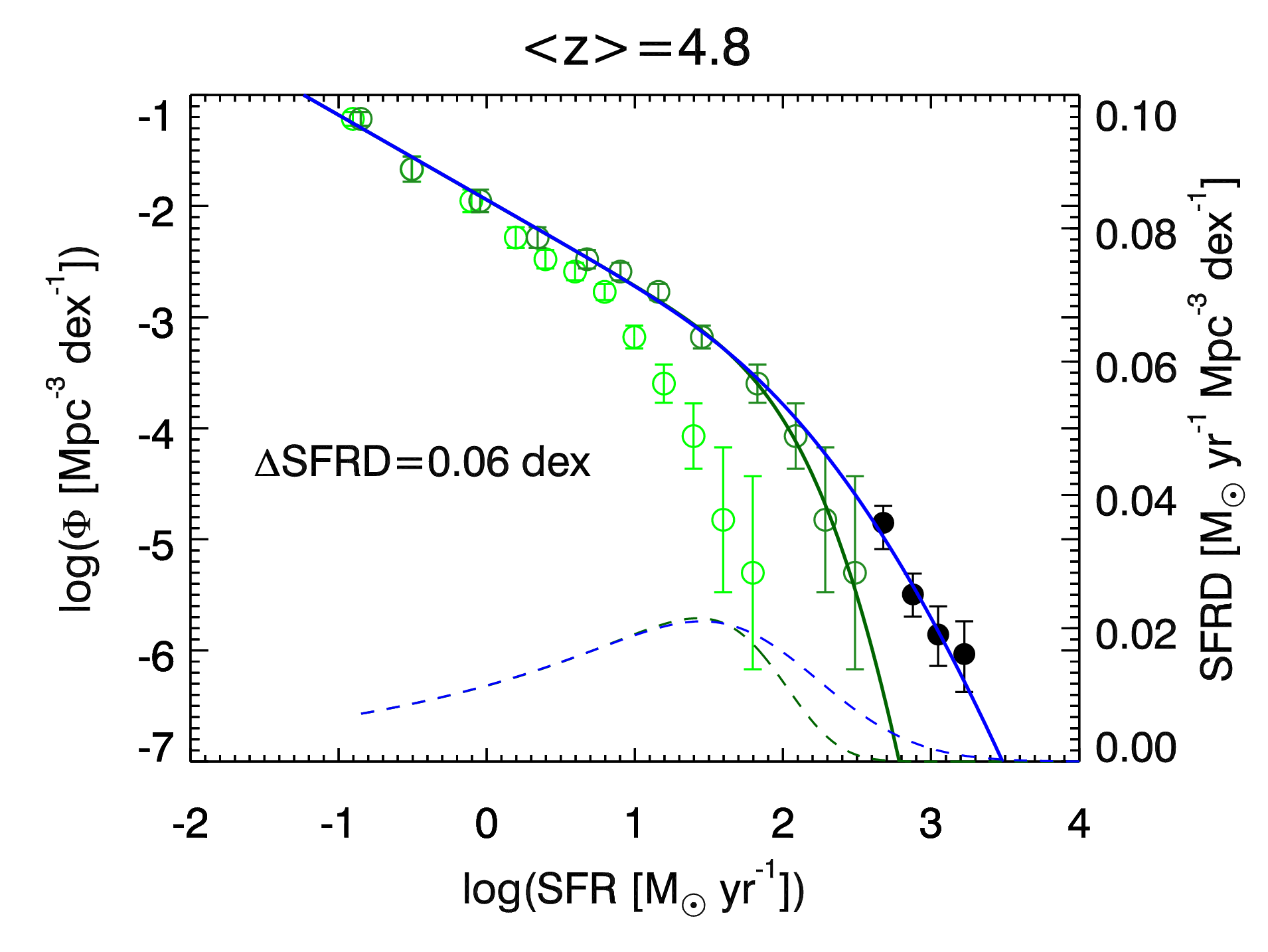}
\caption{ Number density of UV \citep{bouwens15} and our radio SF galaxies as a function of SFR in the two highest redshift bins. Dust-corrected (uncorrected) UV data are shown with dark (light) green open circles, and our radio data are shown with filled black circles. A fit with the functional form given in Eq.~(\ref{eq:lflocal}) is performed on the UV data only (green full line) and the radio plus faint UV data (blue full line). Dashed lines show the SFRD contribution with the scale given on the right axis. See text for details.}
\label{fig:lumfun_uv}
\end{figure}

To simultaneously model both the faint and bright ends of the SF LFs at high redshifts in an attempt to better constrain the SFRD of that epoch we use the  dust-corrected UV LFs from \cite{bouwens15} along with our own radio LFs and perform a fit on the combined data as explained below. 
The UV dust corrections are more severe at high luminosities and the LBG selection criteria can easily miss the most massive and dusty galaxies with significant SFRs. On the other hand, the radio emission is an excellent tracer of such SF galaxies. Therefore, we disregard the three most luminous UV LF points at redshifts $z\sim4$ and $z\sim5$ and fit an analytical form given in Eq.~(\ref{eq:lflocal}) to the remaining UV points combined with all of our radio LFs at the same redshift. The combined data span more than four decades in luminosities.
Our results are shown in Fig.~\ref{fig:lumfun_uv}, where we show the SFR on the x-axis instead of the usual luminosity. Ultraviolet luminosities were scaled to SFR according to \cite{kennicutt98} relation, while our radio luminosities were scaled using the redshift-dependent \qtir\ given in Eq.~(\ref{eq:qtirz}).
It is not our intention to obtain the best SF LF at these redshifts, but rather to calculate an estimate of the missed SFRD in the LBG sample from the radio perspective. Still, for completeness we report here the best-fit parameters obtained. They are  $\Phi_\star=9.35\times10^{-3}~\text{Mpc}^{-3}\text{dex}^{-1}$, $L_\star=1.81\times10^{22}~\whz$, $\alpha=1.62$, and $\sigma=0.83$ at $z\sim4$
 and 
 $\Phi_\star=1.23\times10^{-3}~\text{Mpc}^{-3}\text{dex}^{-1}$, $L_\star=1.26\times10^{23}~\whz$, $\alpha=1.76$, and $\sigma=0.67$ at $z\sim5$.

The SFRD integral of the best LF fit of the combined dust-corrected UV and radio data is 0.08~dex higher at $z\sim4$ and 0.06~dex higher at $z\sim5$ than the values obtained from the UV data alone in the same luminosity range. These integrated values are also plotted as blue diamonds in panel D of Fig.~\ref{fig:sfrd}.
Assuming the dust corrections calculated by \cite{bouwens14b} are correct and start to become significant only at higher luminosities and SFRs \citep[for details, see also][]{wang&heckman96}, this suggests a 15-20\% underestimation of highly obscured SFR estimated from the rest-frame UV observations. Since our radio LFs are slightly lower than the IR LFs at $z\sim4$ (see Fig.~\ref{fig:grid}), this underestimation could be considered a lower limit.
Also, our pure radio SFRD estimate is likely underestimated at $z\sim4$ due to a rather flat faint end slope, while at $z\sim5$ it is actually higher than the combined UV plus radio estimate owing to a higher normalization of the pure evolution fit. 

Our radio LFs are in very good agreement with the work done by \cite{mancuso16}.
These authors used the continuity equation approach with the main sequence star formation timescales to conclude that the number density of SF galaxies at high redshifts ($z\lesssim7$) cannot be reliably estimated from the UV-data alone, even when corrected for dust extinction. Their results also imply the existence of a high-redshift heavily dust-obscured galaxy population with SFRs larger than $100~\msolyr$.

In their work, \cite{burgarella13} attempted to constrain the SFRD by taking into account dust obscuration using combined IR and UV LFs reported in \cite{gruppioni13} and \cite{cucciati12}, respectively. We show their results as magenta crosses in panel D of Fig.~\ref{fig:sfrd}. It is interesting to note a good agreement in SFRD at $z\simeq4$ between substantially different approaches such as the pure UV-based data, IR plus UV data, and the radio plus UV data. They are all consistent within $\sim$20\%, but at the same time higher than previously reported SFRD fits \citep{madau14, behroozi13}.
Work carried out by \cite{dunlop16} is another example that aims at a complete dust-obscured and unobscured (UV + FIR) SFRD census at high redshifts  utilizing ALMA observations of the Hubble Ultra Deep Field (HUDF) at 1.3~mm.  These investigators estimate UV contribution to the total SFR from evolving luminosity functions given in \cite{parsa16} and \cite{bouwens15}. \cite{dunlop16} find SFRD (shown as red squares in panel D of Fig.~\ref{fig:sfrd}) consistent with \cite{behroozi13} in the redshift range $2.5<z<4.5$. They also find a transition in the dominant SF population from dust obscured to dust unobscured at $z\gtrsim4$.
Given the wide distribution and uncertainties of calculated SFRD arising in insufficient knowledge of dust corrections, we believe that the inclusion of radio observations as a dust-unbiased tracer can help achieve a better consensus.

\subsection{ULIRGs and HyLIRGs}

In panel E of Fig.~\ref{fig:sfrd} we decompose our total SFRD to focus on galaxies that form stars efficiently (SFR $>100~\msolyr$). Our SFRD estimates for ULIRGs and HyLIRGs are shown as purple asterisks and red crosses,  respectively. As previously, the first consistency check is to compare our SFRD for ULIRGs with those estimated by \cite{smolcic09a}, shown as blue downward triangles. 
Our values are slightly higher than theirs, as was the case with the LFs, in the redshift range sampled by \cite{smolcic09a}.
Ultraluminous IR galaxies are well constrained by our data in the redshift range $0.6<z<3.3$, with little to no extrapolation needed.
The data show that ULIRGs contribute above 16\% to the total SFRD at $z>1$ with a peak of $\sim$25\% at a redshift of $z\sim2.5$, while HyLIRGs contribute an additional $\lesssim 2\%$ in the entire observed range.

\cite{caputi07} inferred the bolometric IR ($5-1000~\mu$m) luminosity density up to $z\sim2$ using \textit{Spitzer} 24~$\mu$m selected galaxies in the GOODS fields. We show their SFRD results for ULIRGs only as green dot-dashed line. The agreement is good up to $z\sim1$, but it gets worse at higher redshifts where their estimates are significantly higher than ours. Discrepancies may be caused by a different star-forming galaxy sample selection as mentioned in the previous section where we compared IR- and radio-based SFRD.
Additionally, the luminosity integration limits needed to calculate contributions from ULIRGs are directly scaled (see Sect.~\ref{sec:sfrd_ranges}) by the redshift-dependent \qtir\ parameter. The total scaling effect of the \qtir$(z)$ on the SFRD integral is further discussed in \ref{sec:sys_irr}.

Additionally, in the same panel, we show SFRD based on $\sim 100$ ALMA LESS (ALESS\footnote{LABOCA Extended Chandra Deep Field South (ECDFS) Submm Survey (LESS).}) submillimeter galaxies ($S_{870\mu\text{m}}>1$~mJy)  by \cite{swinbank14} as the magenta dashed line and $1\sigma$ errors as dotted lines. These SMGs are highly dust obscured and have large SFR. 
Since our lower integration limit for ULIRGs ($100~\msolyr$) is slightly higher than theirs ($80~\msolyr$), 0.11~dex should be added to our ULIRG SFRD values prior to comparison with \cite{swinbank14} results. There is a broad agreement within the error bars between these two results.
However, there are some additional complications in comparing these results because their observations are less sensitive to hotter than average dust temperatures, and they report up to a factor of 2 uncertainty due to missing these ULIRGs. Therefore, their results represent lower limits. Also, both the \cite{swinbank14} results and our results rely on non-negligible extrapolations to fainter flux densities.

\section{Potential biases}
\label{sec:sys}

Here we summarize some critical assumptions and associated possible systematic effects on our results. While the biggest uncertainties arise from extrapolations, there are a number of additional redshift-dependent and independent effects that may scale our LFs and SFRD history.

\
\subsection{AGN contamination}
\label{sec:sys_agn}
In this paper we adopted an IR-radio-based discrimination of galaxy populations since our goal was to estimate LFs of star-forming galaxies and the total SFRD history from the radio perspective. We assume that the IR is a good tracer of SFR in our radio detected galaxies and that SF galaxies follow a IR-radio correlation with some intrinsic scatter. Because the observed scatter is nonsymmetrical, i.e., there is a tail of sources with large radio fluxes compared to IR measurements, we conclude that AGN contribution to the radio emission is large in such galaxies. The radio-excess method described in Sect.~\ref{sec:excess} is therefore good at selecting galaxies dominated by AGN emission in the radio band. The $3\sigma$ cut given in Eq.~(\ref{eq:excess_cut}) ensures that only $\sim$0.15\% of removed galaxies are SF, giving us a high level of completeness of our SF sample. On the other hand, by counting sources below the $3\sigma$ radio-excess limit and above the best-fitting symmetric profile in all redshift bins (see Sect.~\ref{sec:excess}) we estimate that the integrated radio emission can be contaminated by some AGN contribution for around 1\,000 sources (17\% of the sample), and this contribution is limited to a maximum of 80\%. This potential AGN contribution is mitigated when calculating the SFRD integral by using a properly calibrated \qtir\ relation (see Sect.~\ref{sec:sys_irr}).
When AGN enter the sample, they increase the density in the LF, but at the same time lower the \qtir\ parameter \citep[see][]{delhaize17}.
If a smaller, less than $3\sigma$, cut were used, then more and more SF galaxies would be removed from the sample trading completeness for purity.

In an attempt to obtain a clean SF sample, free of AGN hosts, we employed a different selection method   explained in \cite{smolcic17b}. We start from the full radio sample with optical-NIR counterparts. The first step in removing AGN includes the use of a cutoff in the X-ray full band (rest-frame $0.5-8$~keV) luminosity \citep[see][]{szokoly04}. In the second step, a warm dusty torus signature around the supermassive black hole  is found in the MIR using a cut in the four IRAC bands as prescribed in \cite{donley12}. The third step uses SED fits with AGN templates \citep{dacunha08,berta13} to exclude galaxies with significant AGN contribution \citep[see also][]{delvecchio17}. These three criteria remove moderate-to-high radiative luminosity AGN from the sample. The next step uses rest-frame optical colors $M_{NUV}-M_r$ corrected for internal dust extinction to select red quiescent galaxies ($M_{NUV}-M_r>3.5$, \citealt{ilbert10}) that may host an AGN detected in the radio. If galaxies with such colors do not have a $5\sigma$ detection in the \textit{Herschel} image, they are then classified as  low-to-moderate radiative luminosity AGN and excluded from our sample. The remaining 4555 green and blue galaxies ($M_{NUV}-M_r \le 3.5$) without a $3\sigma$ radio excess (see Sect.~\ref{sec:excess}) are considered a clean sample of SF galaxies based on available AGN diagnostics. Since this sample does not account for the star formation in AGN hosts, it represents a conservative lower limit to SF LFs and SFRD. When the analysis is repeated with this clean SF sample, we find a median decrease of 0.12~dex in number densities across all observed epochs without significant redshift trends. For consistency, the \qtir\ was recalculated for the clean SF galaxy sample. It gives slightly higher values at all redshifts, but agrees with that given in Eq.~(\ref{eq:qtirz}) within  $1.5\sigma$.
The total SFRD integral median decrease is 0.035~dex, which is within the uncertainties of our nominal sample. The SFRD median decrease is not as significant as the LF median density decrease because we are still fitting the pure luminosity evolution to newly derived LFs, meaning that the faint end remains mostly unchanged; we also recalibrated the \qtir\ parameter to match the clean SF sample.

\subsection{The choice of the local LF}
\label{sec:sys_locallf}
\begin{figure}
\centering
\includegraphics[width=\factor\columnwidth]{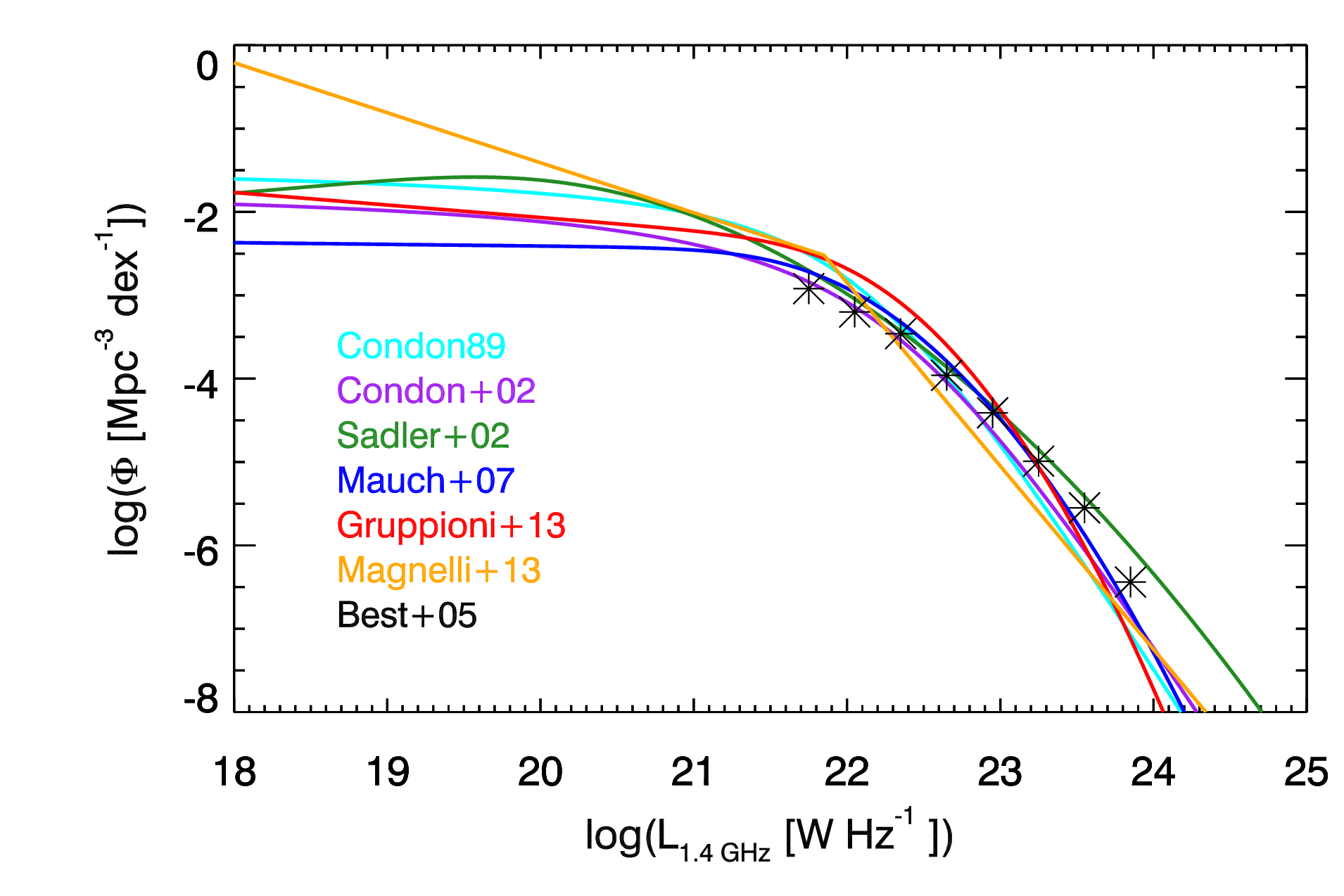}

\includegraphics[width=\factor\columnwidth]{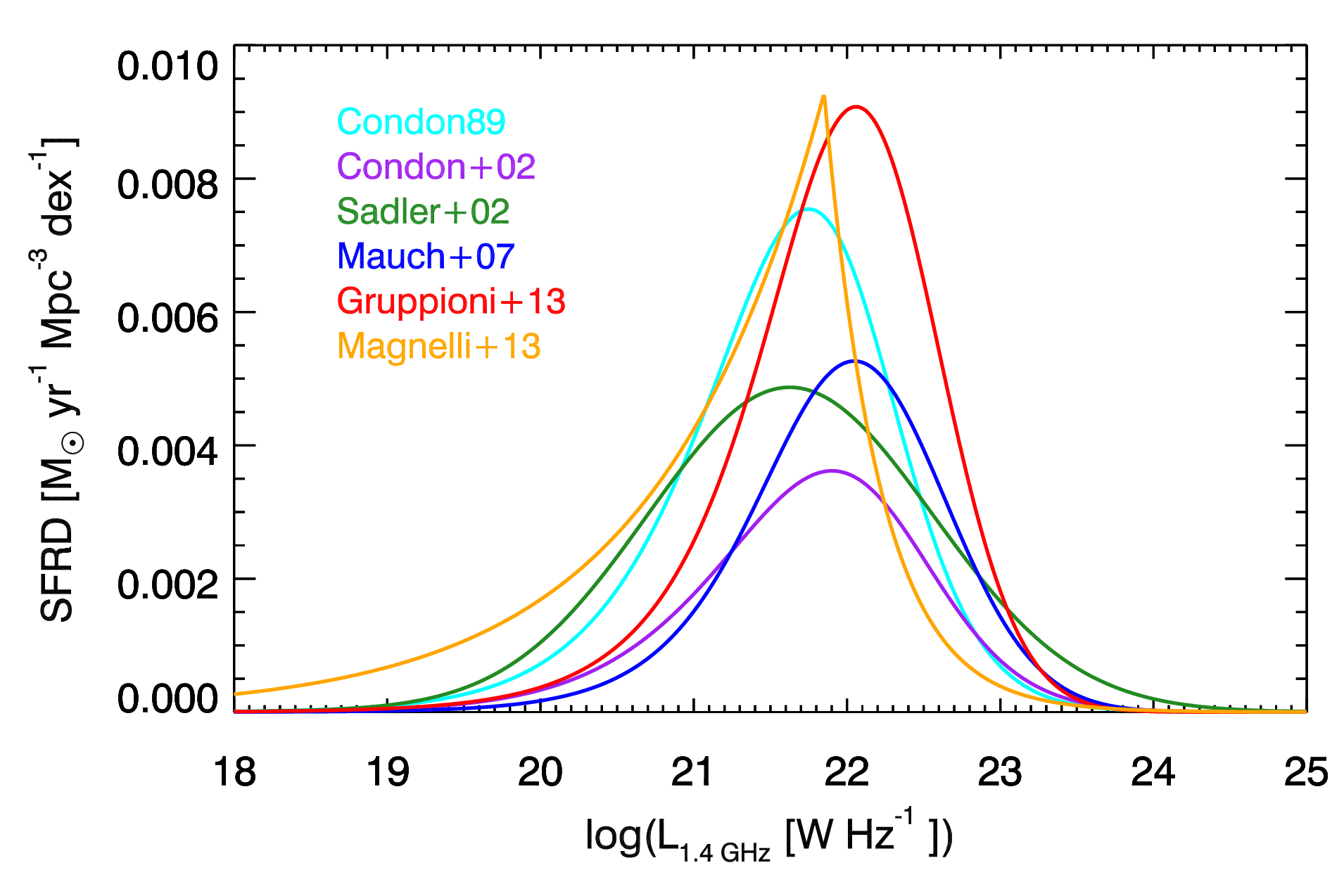}
\caption{\textit{Top}: Local radio and IR LF at 1.4~GHz from various authors as indicated in the legend. Red and orange lines correspond to IR data while all other lines are derived from radio data. \cite{best05} did not attempt to fit an analytical form so we show their points as asterisks. Functional forms are either broken power law (orange), hyperbolic form (cyan and purple), or power law plus lognormal (green, blue, and red). \textit{Bottom}: LFs converted to SFRD per logarithm of luminosity using Eq.~(\ref{eq:sfr}) and a local \qtir=2.64 value from \cite{bell03}.}
\label{fig:local}
\end{figure}

The choice of the analytical shape of the LF can have a significant effect on the total SFRD due to extrapolation toward unobserved luminosities.
A compilation of local LF is shown in the top panel Fig.~\ref{fig:local}. Where necessary, according to \cite{ascasibar02}, LF was corrected for the change of cosmology by scaling $L(z)\propto d_m^2(z) \propto h^{-2}$ and $\Phi(z)\propto d_m^{-3}(z) \propto h^{3}$. 
The bottom panel of Fig.~\ref{fig:local} shows the contribution to the SFRDs as a function of luminosity for the various LFs; although all the LFs show a peak at a similar radio luminosity, the positions of these peaks can differ by up to $\sim$0.3~dex.
There is also no physical argument for the shape of the LF being fix and evolving in redshift by simple translation. However, our data cannot constrain the full luminosity range required to obtain the most significant bulk of the SFRD integral at all redshifts, so this way of extrapolation was chosen for its simplicity.
For a more complex handling of the LF evolution, see for example \cite{fotopoulou16}, where they used a Bayesian approach to model and constrain the shape of the AGN LF as a function of redshift.

\subsection{IR-radio correlation}
\label{sec:sys_irr}

The most significant factor in our SFRD estimates is the \qtir\ parameter since it directly scales our integrated radio luminosities as a function of redshift. Throughout this work we used \qtir$(z)$ estimated on the same SF galaxy sample with the methods from \cite{delhaize17}.
The following are a few underlying assumptions when using such an evolving \qtir(z) value:
\begin{enumerate}
\item The IR emission is an accurate tracer of SFR at all redshifts and radio emission originates mostly in SF processes. 
Extragalactic radio observations can properly trace emission from SF processes in a galaxy  when cosmic ray electrons are not allowed to escape it.
The escape scenario is possible for small sized galaxies with $\lum\lesssim2\times10^{21}~\whz$ \citep[e.g.,][]{bell03}, which is far below our observational limit.
However, the nonthermal radio emission needs a proxy to derive the actual SFR value and the assumption is that the IR emission is a good proxy.\\
\item Infrared-radio correlation is linear, meaning that it can be represented as a single line with a slope of one in the log-log plot of radio and IR luminosities.\\
\item Radio spectrum is a simple power law in frequency. This is a widely used approximation and is often taken for granted  because of insufficient radio data, however, it plays an important role, especially at high redshifts.
\end{enumerate}
Within the framework of these assumptions it is correct to use an evolving \qtir(z) when calculating the SFR of a galaxy from radio emission. 
Even if the second or the third assumption was not correct,  for example, because of various free-free contributions in the radio spectrum or the luminosity dependence of the IR-radio correlation, which might cause a difference between the IR and radio LF evolution, the \qtir$(z)$ evolution takes these wrong assumptions into account and produces a correct SFR value on average because it was calibrated using both the radio and IR data.

To demonstrate the scaling effect of the \qtir\ parameter on our SFRDs we integrate our continuous simple evolution model from Sect.~\ref{sec:simple_model} and show the results with a blue line in panel F of Fig.~\ref{fig:sfrd}, while the shaded area corresponds to the $1\sigma$ uncertainty owing to the errors on the fit parameters added in quadrature with the \qtir$(z)$ uncertainty.
If we instead take the standard constant local value of $\qtir=2.64$ from \cite{bell03} and apply it to our simple LF evolution model, we would obtain three times larger SFRD estimates at  $z\sim4$ (see gray dotted line in the same panel). Observations however do not favor this choice.
Another analysis of the IR-radio correlation was performed through stacking by \cite{magnelli15}. They obtained $q_\text{FIR}(z)=2.35 \times (1+z)^{-0.12}$. This relation can be scaled as $\log(L_\text{FIR})=\log(L_\text{TIR}) - \log(2)$ to obtain the \qtir(z) needed for our conversion, which is valid in terms of median statistics; see also \cite{delhaize17}. The SFRD obtained from this expression is shown as a red dot-dashed line in the same panel and is similar to ours.
To summarize, the trend in the cosmic SFRD history that we obtain from our simple LF evolution model is linked with the trend in the \qtir\ and it is important for this value to be well constrained at all observed redshifts.

\subsection{Radio spectral indices}
\label{sec:sys_alpha}
Regarding the accuracy of the computed rest-frame 1.4~GHz luminosity, the highest uncertainty, especially at high redshifts, lies in the insufficient knowledge of the radio K correction. For example, a rather large photometric error of $\Delta z=0.3$ would result in a 0.05~dex error in luminosity at $z\sim5$. However, an uncertainty in spectral index of just $\Delta \alpha=0.1$ would produce an error of 0.1~dex in luminosity. It is known that a symmetric spread of the spectral indices around the mean value for SF galaxies is approximately $\sim0.4$ \citep[e.g.,][]{kimball08}. 
Even though this spread would produce a significant uncertainty on the estimated \lum\ of each single galaxy, these variations approximately cancel each other and give a valid average luminosity because of the symmetry of the distribution of
spectral indices. 
It is widespread in the literature to assume a single spectral index for radio SED, where the usual values are $\alpha=-0.8$ or $-0.7$.

Approximately 75\% of our radio sources were only detected at 3~GHz.
The use of the measured spectral indices for the remaining 25\% can introduce a small bias toward steeper spectra (therefore higher luminosities), since our survey is currently the deepest radio survey of the COSMOS field. For example, a point source at the limit of our sensitivity ($rms=2.3~\mujybeam$) would have to have a spectral index steeper than -1.9 to be observed in the previous deep 1.4~GHz survey ($rms=10~\mujybeam$; \citealt{schinnerer10}). The median spectral index of sources detected in both surveys is $\alpha=-0.85$.

To assess the impact of the used spectral indices on our results, we repeated the analysis two times: the first time with the standard $\alpha=-0.7$ and the second time with $\alpha=-0.8$ for all sources regardless of the observed radio spectrum.
When a single and identical spectral index is used for each source, the pure luminosity evolution of the chosen analytical local function from Eq.~\ref{eq:lflocal} better fits (smaller $\chi^2$) the derived LFs at all luminosities and redshifts. 
Specifically, when $\alpha=-0.7$ is used, the best pure luminosity fit evolution remains essentially unchanged from that presented in Sect.~\ref{sec:res_lumfun}, which is unsurprising given that 75\% of the spectral indices remained unchanged as well.
When $\alpha=-0.8$ is used, a stronger luminosity evolution is obtained, which is described by an increase of 0.16 in the evolution parameter $\alpha_{L}$ as given in
Eq.~(\ref{eq:lumfun_evol}) and previously shown in Fig.~\ref{fig:knee}. This increase is still within the model uncertainties as given in Sect.~\ref{sec:simple_model}.
Before deriving the SFRD values, for consistency, the \qtir\ parameter was again recalculated using different spectral indices and obtained expressions are within  the $1.5\sigma$ of the nominal values given in Eq.~(\ref{eq:qtirz}). Derived total SFRDs are within the uncertainties of the nominal sample in both cases, which strengthens the robustness of our results.

Finally, assuming a simple power-law radio spectra might be an overly simplistic approach given the unknown contribution of free-free (Bremsstrahlung) emission to the total SED.  Additional deep radio observations at higher frequencies are needed to properly model the radio SED and mitigate this limitation.

\section{Summary}
\label{sec:summary}
We  studied a radio-selected sample of star-forming galaxies from deep VLA-COSMOS 3~GHz observations \citep{smolcic17a}. Galaxy classifications were performed by relying heavily on the optical-NIR COSMOS2015 catalog \citep{laigle16} and SED fits by \cite{delvecchio17}. The final sample contains 5\,915 galaxies, where the radio emission is not dominated by an AGN.
Using this sample we derived radio LFs up to $z\sim5$. By comparing them with LFs derived using IR and UV selected samples we checked their robustness and found that our radio LF can be very well described by a local LF with a power-law plus lognormal form evolved only in luminosity as ${\lum \propto(1+z)^{(\totalfa)-(\totbeta) z}}$. However, we do not observe the faint end of the LF at all redshifts to properly constrain a more complex evolution.
The difference between radio and UV LFs suggests an underestimation of dust corrections obtained from UV slopes by \cite{bouwens14b}.
We converted our radio luminosities to SFR using a redshift-dependent IR-radio correlation where \qtir\ parameter decreases with increasing redshift \citep{delhaize17}. An accurate constraint on this parameter is the most important factor for estimating SFR from radio observations in the early universe.
Our data suggest that the peak of the total SFRD history occurs at $2<z<3$.
We find that the total SFRD estimates using only LBG galaxies \citep[e.g.,][]{bouwens15}, even if corrected for dust extinction, are still likely to miss up to 15-20\% of SFR in highly obscured galaxies at $z\gtrsim 4$. 
By integrating LF fits in various luminosity limits we estimated SFRDs of the total SF sample and the subpopulations of the sample, such as ULIRGs and HyLIRGs.
We find that ULIRGs contribute at maximum up to $\sim$25\% of the total SFRD at $z\sim2.5$, where this population of galaxies is well constrained by our data.
Even though HyLIRGs can have very large SFRs (several 1000~\msolyr), we find that they contribute less than 2\% to the total SFRD at all redshifts owing to their low volume density.

\begin{acknowledgements}
This research was funded by the European Unions Seventh Frame-work program under grant agreement 337595 (ERC Starting Grant, 'CoSMass').
ES warmly acknowledges support from the National Radio Astronomy Observatory for
visits to the Array Operations Center in Socorro, NM, in conjunction with this project.
AK acknowledges support by the Collaborative Research Council 956, sub-project A1, funded by the Deutsche Forschungsgemeinschaft (DFG).
MB and PC acknowledge support from the PRIN-INAF 2014.

\end{acknowledgements}

\bibliographystyle{aa}
\bibliography{bibtex}


\begin{table*}
\begin{center}
\caption{Luminosity functions of star-forming galaxies obtained with the \vmax\ method.}
\renewcommand{\arraystretch}{1.5}
\begin{tabular}[t]{c c c c}
\hline
Redshift
 & $\log\left(\dfrac{L_{1.4\,\text{GHz}}}{\text{W}\,\text{Hz}^{-1}}\right)$
 & $\log\left(\dfrac{\Phi}{\text{Mpc}^{-3}\,\text{dex}^{-1}}\right)$
 & N \\
\hline 
$0.1<z<0.4$ & 21.77$_{-1.1}^{+0.23}$ & -2.85$_{-0.077}^{+0.094}$ & 189 \\
 & 22.15$_{-0.15}^{+0.18}$ & -2.88$_{-0.029}^{+0.031}$ & 217 \\
 & 22.46$_{-0.14}^{+0.19}$ & -3.12$_{-0.034}^{+0.037}$ & 149 \\
 & 22.77$_{-0.12}^{+0.20}$ & -3.55$_{-0.055}^{+0.063}$ & 56 \\
 & 23.09$_{-0.12}^{+0.21}$ & -4.05$_{-0.090}^{+0.11}$ & 19 \\
 & 23.34$_{-0.048}^{+0.28}$ & -4.63$_{-0.22}^{+0.25}$ & 5 \\
$0.4<z<0.6$ & 22.29$_{-0.31}^{+0.11}$ & -2.97$_{-0.046}^{+0.052}$ & 142 \\
 & 22.54$_{-0.14}^{+0.13}$ & -3.19$_{-0.033}^{+0.036}$ & 160 \\
 & 22.80$_{-0.12}^{+0.15}$ & -3.33$_{-0.035}^{+0.038}$ & 143 \\
 & 23.04$_{-0.090}^{+0.18}$ & -3.67$_{-0.052}^{+0.059}$ & 65 \\
 & 23.31$_{-0.096}^{+0.18}$ & -4.32$_{-0.097}^{+0.12}$ & 16 \\
 & 23.68$_{-0.19}^{+0.081}$ & -5.05$_{-0.30}^{+0.34}$ & 3 \\
$0.6<z<0.8$ & 22.61$_{-0.23}^{+0.080}$ & -2.89$_{-0.075}^{+0.091}$ & 179 \\
 & 22.84$_{-0.15}^{+0.15}$ & -3.13$_{-0.025}^{+0.027}$ & 283 \\
 & 23.11$_{-0.12}^{+0.17}$ & -3.47$_{-0.033}^{+0.035}$ & 165 \\
 & 23.40$_{-0.12}^{+0.17}$ & -3.99$_{-0.057}^{+0.066}$ & 51 \\
 & 23.71$_{-0.13}^{+0.16}$ & -4.68$_{-0.11}^{+0.16}$ & 11 \\
 & 24.06$_{-0.19}^{+0.10}$ & -5.43$_{-0.37}^{+0.45}$ & 2 \\
$0.8<z<1.0$ & 22.85$_{-0.17}^{+0.074}$ & -3.01$_{-0.041}^{+0.046}$ & 172 \\
 & 23.05$_{-0.13}^{+0.13}$ & -3.13$_{-0.024}^{+0.025}$ & 312 \\
 & 23.30$_{-0.12}^{+0.14}$ & -3.45$_{-0.030}^{+0.032}$ & 198 \\
 & 23.54$_{-0.099}^{+0.16}$ & -3.85$_{-0.046}^{+0.051}$ & 82 \\
 & 23.81$_{-0.11}^{+0.15}$ & -4.31$_{-0.073}^{+0.088}$ & 30 \\
 & 24.11$_{-0.15}^{+0.11}$ & -4.89$_{-0.17}^{+0.18}$ & 8 \\
$1.0<z<1.3$ & 23.10$_{-0.21}^{+0.081}$ & -3.19$_{-0.046}^{+0.052}$ & 216 \\
 & 23.31$_{-0.12}^{+0.15}$ & -3.42$_{-0.024}^{+0.025}$ & 321 \\
 & 23.57$_{-0.12}^{+0.16}$ & -3.86$_{-0.034}^{+0.036}$ & 156 \\
 & 23.84$_{-0.11}^{+0.16}$ & -4.15$_{-0.046}^{+0.052}$ & 81 \\
 & 24.06$_{-0.051}^{+0.22}$ & -4.74$_{-0.084}^{+0.10}$ & 22 \\
 & 24.38$_{-0.10}^{+0.17}$ & -5.25$_{-0.19}^{+0.20}$ & 7 \\
$1.3<z<1.6$ & 23.32$_{-0.16}^{+0.070}$ & -3.36$_{-0.039}^{+0.043}$ & 156 \\
 & 23.53$_{-0.14}^{+0.18}$ & -3.55$_{-0.024}^{+0.025}$ & 323 \\
 & 23.81$_{-0.10}^{+0.21}$ & -4.10$_{-0.037}^{+0.041}$ & 126 \\
 & 24.15$_{-0.12}^{+0.19}$ & -4.53$_{-0.059}^{+0.068}$ & 48 \\
 & 24.39$_{-0.053}^{+0.26}$ & -5.30$_{-0.17}^{+0.18}$ & 8 \\
 & 24.82$_{-0.17}^{+0.14}$ & -5.94$_{-0.37}^{+0.45}$ & 2 \\

\hline
\end{tabular}
\quad
\renewcommand{\arraystretch}{1.5}
\begin{tabular}[t]{c c c c}
\hline
Redshift
 & $\log\left(\dfrac{L_{1.4\,\text{GHz}}}{\text{W}\,\text{Hz}^{-1}}\right)$
 & $\log\left(\dfrac{\Phi}{\text{Mpc}^{-3}\,\text{dex}^{-1}}\right)$
 & N \\
\hline  
$1.6<z<2.0$ & 23.55$_{-0.18}^{+0.065}$ & -3.47$_{-0.070}^{+0.084}$ & 156 \\
 & 23.72$_{-0.11}^{+0.18}$ & -3.66$_{-0.024}^{+0.025}$ & 312 \\
 & 23.98$_{-0.080}^{+0.21}$ & -4.15$_{-0.035}^{+0.038}$ & 141 \\
 & 24.28$_{-0.10}^{+0.18}$ & -4.56$_{-0.054}^{+0.062}$ & 57 \\
 & 24.53$_{-0.059}^{+0.23}$ & -5.06$_{-0.092}^{+0.12}$ & 18 \\
 & 24.90$_{-0.14}^{+0.14}$ & -5.86$_{-0.30}^{+0.34}$ & 3 \\
$2.0<z<2.5$ & 23.74$_{-0.15}^{+0.086}$ & -3.61$_{-0.045}^{+0.050}$ & 141 \\
 & 23.94$_{-0.11}^{+0.13}$ & -3.86$_{-0.029}^{+0.031}$ & 219 \\
 & 24.19$_{-0.12}^{+0.13}$ & -4.35$_{-0.042}^{+0.046}$ & 98 \\
 & 24.43$_{-0.11}^{+0.13}$ & -4.71$_{-0.061}^{+0.072}$ & 44 \\
 & 24.66$_{-0.10}^{+0.14}$ & -5.04$_{-0.086}^{+0.11}$ & 21 \\
 & 24.96$_{-0.15}^{+0.092}$ & -5.61$_{-0.20}^{+0.22}$ & 6 \\
$2.5<z<3.3$ & 24.01$_{-0.21}^{+0.079}$ & -3.96$_{-0.051}^{+0.057}$ & 128 \\
 & 24.20$_{-0.11}^{+0.13}$ & -4.21$_{-0.034}^{+0.037}$ & 155 \\
 & 24.42$_{-0.091}^{+0.15}$ & -4.62$_{-0.046}^{+0.051}$ & 81 \\
 & 24.68$_{-0.11}^{+0.13}$ & -4.94$_{-0.065}^{+0.076}$ & 39 \\
 & 24.92$_{-0.11}^{+0.13}$ & -5.43$_{-0.12}^{+0.16}$ & 11 \\
 & 25.18$_{-0.13}^{+0.10}$ & -6.27$_{-0.37}^{+0.45}$ & 2 \\
$3.3<z<4.6$ & 24.30$_{-0.24}^{+0.097}$ & -4.58$_{-0.079}^{+0.096}$ & 55 \\
 & 24.48$_{-0.081}^{+0.12}$ & -5.07$_{-0.077}^{+0.093}$ & 27 \\
 & 24.67$_{-0.074}^{+0.13}$ & -5.24$_{-0.082}^{+0.10}$ & 23 \\
 & 24.86$_{-0.066}^{+0.13}$ & -5.76$_{-0.20}^{+0.22}$ & 6 \\
 & 25.13$_{-0.13}^{+0.067}$ & -5.91$_{-0.22}^{+0.25}$ & 5 \\
$4.6<z<5.7$ & 24.51$_{-0.13}^{+0.085}$ & -4.85$_{-0.15}^{+0.24}$ & 11 \\
 & 24.71$_{-0.12}^{+0.095}$ & -5.50$_{-0.19}^{+0.20}$ & 7 \\
 & 24.88$_{-0.076}^{+0.13}$ & -5.86$_{-0.25}^{+0.28}$ & 4 \\
 & 25.06$_{-0.042}^{+0.17}$ & -6.03$_{-0.30}^{+0.34}$ & 3 \\

\hline
\end{tabular}
\label{tab:lumfun_vmax}
\end{center}
\end{table*}

\begin{table*}
\begin{center}
\caption{Cosmic SFR density history obtained by integrating the analytical form of the best-fit LF in different redshift bins. All SFRD estimates except the last column refer to pure luminosity evolution. For a combined density and luminosity evolution only the 68\% confidence interval is reported.}
\renewcommand{\arraystretch}{1.5}
\begin{tabular}[t]{c c c c c c }
\hline

Redshift & Total & Lower limit & ULIRGs & HyLIRGs & $\Phi$ and $L$ evolution \\
z & \multicolumn{5}{c}{$\log(\textrm{SFRD}[\msolyrmpc])$} \\
\hline
0.312$_{-0.21}^{+0.088}$ & -1.77$_{-0.052}^{+0.047}$ & -1.93$_{-0.059}^{+0.052}$ & -3.11$_{-0.11}^{+0.099}$ & -5.05$_{-0.15}^{+0.14}$ & [-1.83, -1.73] \\
0.501$_{-0.10}^{+0.099}$ & -1.69$_{-0.050}^{+0.045}$ & -1.98$_{-0.064}^{+0.057}$ & -2.93$_{-0.098}^{+0.088}$ & -4.78$_{-0.14}^{+0.12}$ & [-1.80, -1.68] \\
0.695$_{-0.095}^{+0.11}$ & -1.47$_{-0.048}^{+0.044}$ & -1.78$_{-0.061}^{+0.055}$ & -2.43$_{-0.082}^{+0.074}$ & -4.07$_{-0.11}^{+0.10}$ & [-1.52, -1.36] \\
0.903$_{-0.10}^{+0.097}$ & -1.31$_{-0.048}^{+0.044}$ & -1.64$_{-0.058}^{+0.053}$ & -2.09$_{-0.072}^{+0.065}$ & -3.56$_{-0.099}^{+0.091}$ & [-1.38, -1.27] \\
1.16$_{-0.16}^{+0.14}$ & -1.31$_{-0.049}^{+0.043}$ & -1.75$_{-0.062}^{+0.054}$ & -2.10$_{-0.070}^{+0.060}$ & -3.56$_{-0.095}^{+0.082}$ & [-1.44, -1.28] \\
1.44$_{-0.14}^{+0.16}$ & -1.24$_{-0.051}^{+0.045}$ & -1.76$_{-0.066}^{+0.056}$ & -1.97$_{-0.069}^{+0.059}$ & -3.38$_{-0.094}^{+0.080}$ & [-1.39, -1.21] \\
1.81$_{-0.21}^{+0.19}$ & -1.16$_{-0.053}^{+0.047}$ & -1.74$_{-0.068}^{+0.059}$ & -1.81$_{-0.068}^{+0.060}$ & -3.12$_{-0.091}^{+0.080}$ & [-1.38, -1.15] \\
2.18$_{-0.18}^{+0.32}$ & -1.10$_{-0.056}^{+0.048}$ & -1.77$_{-0.073}^{+0.061}$ & -1.70$_{-0.071}^{+0.059}$ & -2.96$_{-0.094}^{+0.077}$ & [-1.37, -1.13] \\
2.81$_{-0.31}^{+0.49}$ & -1.08$_{-0.060}^{+0.052}$ & -1.95$_{-0.083}^{+0.071}$ & -1.68$_{-0.076}^{+0.065}$ & -2.92$_{-0.10}^{+0.087}$ & [-1.53, -1.26] \\
3.71$_{-0.41}^{+0.89}$ & -1.23$_{-0.079}^{+0.062}$ & -2.57$_{-0.15}^{+0.11}$ & -1.95$_{-0.12}^{+0.088}$ & -3.34$_{-0.17}^{+0.12}$ & [-2.12, -1.14] \\
4.83$_{-0.23}^{+0.87}$ & -1.25$_{-0.13}^{+0.085}$ & -2.86$_{-0.32}^{+0.18}$ & -1.98$_{-0.24}^{+0.14}$ & -3.39$_{-0.36}^{+0.21}$ & [-2.52, -0.142] \\

\hline
\end{tabular}
\label{tab:lumfun_sfrd}
\end{center}
\end{table*}

\end{document}